\documentclass[journal]{IEEEtran}
\usepackage[utf8]{inputenc}
\usepackage[T1]{fontenc}

\usepackage{graphicx}
\usepackage{subfigure}
\usepackage{caption}
\usepackage{nth}
\usepackage{soul}
\usepackage{float}
\usepackage{xurl}
\usepackage{adjustbox}
\usepackage{multirow}
\usepackage{listings}
\usepackage{paralist}
\usepackage{ragged2e}

\usepackage[inline, shortlabels]{enumitem}
\usepackage[noadjust]{cite}
\usepackage[normalem]{ulem}
\usepackage[inline]{enumitem}

\usepackage{algorithm}
\usepackage{algpseudocode}
\usepackage{amsmath}

\usepackage[hidelinks]{hyperref}
\hypersetup{colorlinks=true,linkcolor=blue,citecolor=blue,urlcolor=blue}

\usepackage{amssymb, amsmath, amsfonts, mathtools, optidef, amsthm} 
\usepackage{blkarray, bigstrut, gauss}

\usepackage{fontawesome5}

\usepackage{siunitx}

\usepackage{array, tabularx, makecell, multirow, booktabs}
\setcellgapes{2.5pt}
\newcolumntype{L}[1]{>{\raggedright\arraybackslash}p{#1}}
\newcolumntype{Z}{>{\centering\let\newline\\\arraybackslash\hspace{0pt}}X}
\newcolumntype{Y}{>{\raggedright\let\newline\\\arraybackslash\hspace{0pt}}X}

\usepackage{siunitx}
\theoremstyle{definition} 
\newtheorem{definition}{Definition}
\newtheorem{example}{Example}

\usepackage{color, xcolor}
\newcommand{\sghaffar}{\color{black}}
\newcommand{\sepehrgh}[1]{\textcolor{black}{#1}}
\newcommand{\sepehr}[1]{\textcolor{black}{#1}}

\newcommand{\subhead}[1]{\noindent{\textbf{#1:}}} 

\usepackage{pifont}

\usepackage{xspace}

\newcommand{\eg}{\textit{e.g.,}\xspace}
\newcommand{\ie}{\textit{i.e.}\xspace}
\newcommand{\etal}{\textit{et al.}\xspace}
\newcommand{\mitre}{\textsc{MITRE ATT\&CK}\xspace}
\newcommand{\sol}{\textsc{AutoSigma}\xspace}

\usepackage[compact]{titlesec}
\titlespacing{\section}{.6pt}{*.6}{*.6}
\titlespacing{\subsection}{0.6pt}{*0.6}{*0.6}
\titlespacing{\subsubsection}{0.5pt}{*0.5}{*0.5}

\usepackage{xcolor}
\usepackage{tikz}
\definecolor{autosigmaYellow}{HTML}{FFA33C}

\newcommand{\squircle}[1]{%
  \tikz[baseline=(char.base)]{
    \node[
      draw=none, 
      fill=autosigmaYellow,
      minimum width=14pt, 
      minimum height=14pt,
      rounded corners=8pt,
      text=white, 
      font=\bfseries, 
      inner sep=1.2pt
    ] (char) {#1};
  }
}

\definecolor{rqtagcolor}{RGB}{70,117,188}

\usepackage[most]{tcolorbox}
\newlength{\rqtagwidth}
\newlength{\rqtaghalf}

\newcommand{\navycircle}[1]{%
    \tikz[baseline=(char.base)]{
        \node[shape=circle, draw=none, fill=blue!80!black, text=white, 
        font=\bfseries\footnotesize, inner sep=1.2pt] (char) {#1};
    }
}

\tcbset{
  rqbox/.style={
    enhanced, breakable, sharp corners,
    colback=white, colframe=black!35, boxrule=0.5pt,
    arc=1.2mm, left=2mm, right=2mm, top=1.2mm, bottom=1.2mm
  }
}

\newtcolorbox[auto counter]{RQ}{rqbox,
  before upper={\textsc{\bf RQ\thetcbcounter.}\ }
}

\newtcolorbox{RQ*}[1]{rqbox,
  before upper={\textsc{\bf RQ#1.}\ }
}

\begin{document}
\bstctlcite{IEEEexample:BSTcontrol}

\title{
    From Threat Intelligence to Detection: 
    Knowledge-driven Enrichment and Template-based Rule Grounding for Automated Sigma Rule Generation
}

\author{
    \IEEEauthorblockN{
        Sepehr Ghaffarzadegan,
        Boubakr Nour,
        Makan Pourzandi,
        Mourad Debbabi, and 
        Chadi Assi
    }   
    \thanks{This work was made possible in part through the support of the National Cybersecurity Consortium and the Government of Canada. This work was also supported in part by Ericsson Research and the Security Research Centre of Concordia University.}

    \thanks{S. Ghaffarzadegan, M. Debbabi, and C. Assi are with Concordia University, Montréal, Canada (email: first.last@concordia.ca)} 
  
  \thanks{B. Nour and M. Pourzandi are with Ericsson Research, Ericsson, Montréal, Canada (email: first.last@ericsson.com).}
}
    
\markboth{}{}

\maketitle

\begin{abstract}
    Mechanisms for dynamically converting cyber threat intelligence (CTI) into actionable detection capabilities are necessary due to the \textcolor{black}{rapid evolution of Advanced Persistent Threats (APTs)}. Sigma rules are an essential part of contemporary threat detection workflows because they offer a platform-independent framework for expressing detection logic that can be converted into particular queries across SIEM systems. Conventional techniques for manually crafting Sigma rules are prone to mistakes, and necessitate extensive knowledge, which restricts their scalability. Although there are open-source and industry-maintained Sigma rule repositories, they often fail to keep pace with emerging threats and require frequent customization to fit diverse operational environments. This emphasizes the necessity of dynamic rule generation that is adapted to evolving attack techniques as well as particular use cases. In this work, we design \sol, an automated solution for transforming unstructured CTI reports into relevant Sigma rules. Rather than relying solely on language models, \sol leverages a structured knowledge base to enrich partial inputs, matches the enriched content against a repository of existing Sigma rules, and then employs an LLM-as-a-Judge mechanism to iteratively validate the rules. By combining knowledge-driven enrichment, template-based rule grounding, and a multi-stage solution, \sol enables accurate, context-aware, and relevant rule generation. {Evaluations across multiple real-world APT reports and multiple security blogs demonstrate that \sol outperforms alternative solutions and LLM models in rule validity, rule relevancy, \mitre technique coverage, and robustness to input quality.}
    
    \sepehrgh{\small \normalfont \noindent \faYoutube~\sol's Demo: \url{https://youtu.be/iSr6IurQ6BM}}
\end{abstract}


\section{Introduction}
\label{sec:introduction}
Advanced Persistent Threats (APTs) represent a growing menace to
organizations and critical infrastructure. In practice, defenders rely
heavily on detailed Cyber Threat Intelligence (CTI) reports from
industry leaders (\eg Mandiant~\cite{mandiant2024}, CrowdStrike~\cite{crowdstrike2024threatreport}). These reports
document the tactics, techniques, procedures (TTPs), malware, and
infrastructure used by attackers, providing invaluable context for
defense. Despite this wealth of information, CTI reports are
typically published as unstructured narratives rather than
machine-readable or actionable outputs ready-to-use by practitioners. The widespread use of detection rules in industry further highlights the need for adaptable and relevant detection logic. In practice, security teams often need to write new detection rules (\eg Sigma) manually to support proactive threat hunting
and to address diverse operational contexts. Translating these unstructured threat descriptions into
concrete detection content (\eg Sigma rules) is thus a bottleneck in a proactive defense stance when the Sigma rules can be
generated at run-time from internal and external threat intelligence
reports~\cite{suominen2024cyber}. Existing CTI reports often emphasize known indicators of
compromise (IoCs)  
which can be useful for detection but have limited longevity. To counter evolving threats, analysts must extract deeper patterns (\eg attack sequences or
behaviors) from narrative text beyond transient indicators. The Sigma format has become a common
schema for writing generic detection rules, therefore a good candidate to extract these deeper patterns. At the
same time, crafting Sigma rules manually is labor-intensive and
error-prone. Automating this pipeline from raw CTI text to relevant
Sigma rules remains an open challenge.

Prior efforts have tackled parts of this problem. For example, \textsc{TTPDrill}~\cite{husari2017ttpdrill} uses an ontology-based approach to parse CTI text, automatically extracting threat actions and mapping them to \mitre techniques. \textsc{ThreatRaptor}~\cite{gao2021enabling} builds on this idea by linking extracted indicators and actions into a threat behavior graph that can be searched against system logs. Similarly, \textsc{LADDER}~\cite{alam2024ladder} aggregates attack patterns across multiple CTI sources to capture multi-stage intrusions. These systems demonstrate the value of structuring CTI, but they fall behind in producing ready-to-deploy detection rules. They often require well-structured input or significant manual curation, and they do not integrate with a rule repository to ground outputs in known detection logic.
More recently, large language models (LLMs) have been applied to accelerate CTI to rule generation. Work in~\cite{zhang2024llms} and~\cite{ferrag2024generativeailargelanguage} review how LLMs and generative AI are transforming cybersecurity, including threat intelligence extraction and rule generation. These studies suggest LLMs can effectively interpret natural-language reports, but also highlight that automated rule synthesis remains immature.

\begin{figure}[!h]
    \centering
    \includegraphics[width=\linewidth]{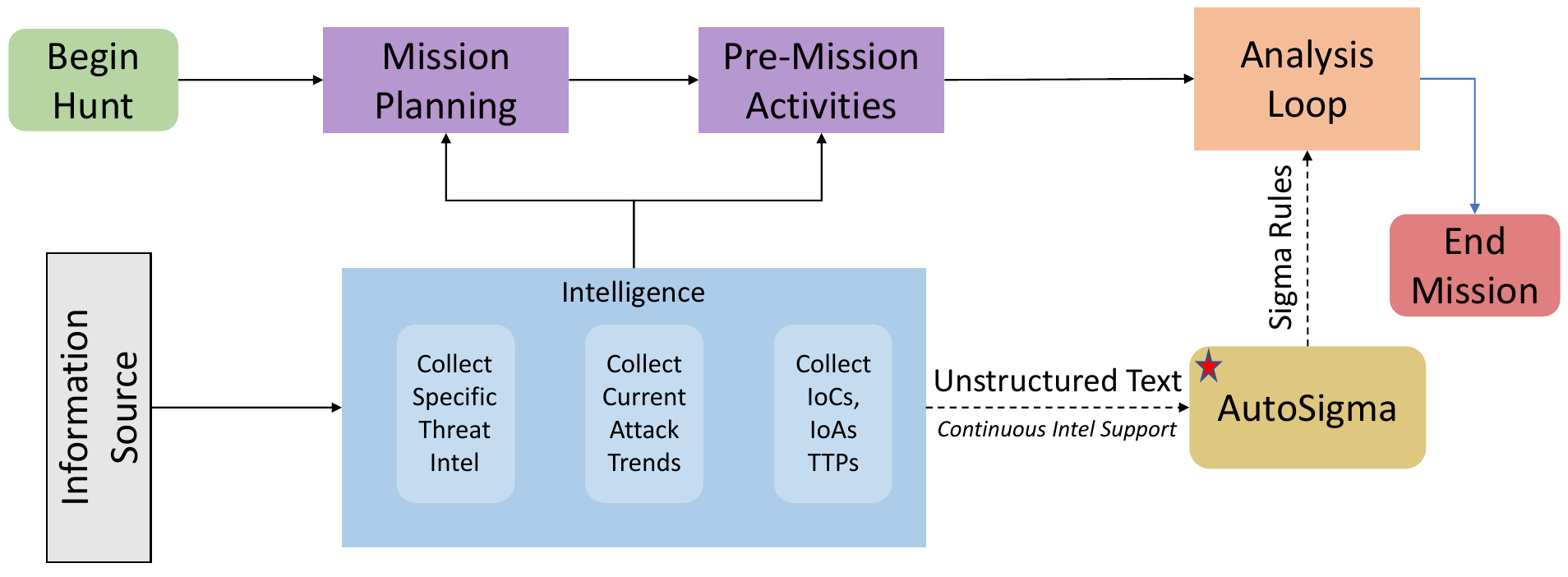}
    \caption{Threat hunting process~\cite{wiliam2024interview} with emphasis on our work focus ($\textcolor{red}{\bigstar}$).}
    \label{fig:threat-hunting-loop}
\end{figure}

\subhead{Our Contribution}
Fig.~\ref{fig:threat-hunting-loop} illustrates how \sol helps and enhances the threat hunting process~\cite{wiliam2024interview}. CTI reports, attack trends, and behavioral indicators drive analysis and mission planning. \sol automates a core bottleneck in this loop by transforming narrative threat intelligence into structured Sigma rules. 

\sepehrgh{Although recent work has made progress toward automating certain CTI processing tasks, there is still no end-to-end solution capable of taking raw narrative reports and producing validated Sigma rules, aligned with \mitre and applicable to different attack types, whether living-off-the-land, cloud-based, or otherwise.}
To address this gap, we present \sol, an end-to-end solution that transforms unstructured CTI reports into precise, deployable Sigma rules.
 Our contributions can be summarized as follows:
\begin{itemize}[left=1pt]
    \item \textit{Fully automated pipeline:} Unlike prior work~\cite{10.1145/3696410.3714798, xu2024intelexllmdrivenattacklevelthreat}, \sol performs contextual analysis, template-driven rule retrieval, and generative refinement in a unified and automated pipeline that helps in reducing manual effort.
    
    \item \textit{Attack scenario enrichment:} \sol enhances attack understanding by using external cybersecurity knowledge bases to fill in missing details and extend raw threat descriptions, which enables robust rule generation even with sparse input.
    
    \item \textit{Attack decomposition:} \sol decomposes complex attack scenarios into smaller, atomic steps aligned with known detection logic. This improves the accuracy, granularity, and modularity of the generated Sigma rules. 
    
    \item \textit{Template-based rule grounding:} By retrieving similar Sigma rules from validated repositories, \sol uses them as templates to guide and constrain rule generation, which ensures syntactic correctness and contextual alignment with industry best practices.
    
    \item \textit{LLM-as-a-Judge refinement:} \sol incorporates a dual-LLM feedback mechanism that iteratively validates and improves candidate rules, which helps in mitigating hallucination, enhancing quality, and ensuring rule validity.
    
    \item \textit{Real-world validation:} We empirically evaluate \sol on \sepehrgh{multiple public security blogs and also APT reports (APT41, APT28, and APT29) against state-of-the-art solution and LLM models}, demonstrating high-fidelity rule generation, accurate threat behavior capture, and readiness for deployment in operational SIEM environments.
\end{itemize}

\subhead{Advantages of \sol}
With its LLM-driven refinement mechanism, \sol offers several advantages over prior approaches. 

First, unlike \textsc{LLMCloudHunter}~\cite{10.1145/3696410.3714798} that requires highly detailed reports, \sol can compensate for incomplete intelligence by enriching attack descriptions with external knowledge bases,
and thus accelerate rule generation and provide more threat coverage.

Second, unlike traditional single-pass generation or standard self-refinement methods~\cite{10.1145/3696410.3714798}, \sol introduces a feedback refinement loop driven by a dual LLM-as-a-Judge that aims to generate rules and refine them for better accuracy. This approach ensures consistent formatting and syntax of Sigma rules, which reduces human errors and improves detection accuracy.

Finally, \sol eases the operationalization of threat intelligence to be used in security operations without requiring expert rule engineers for every new threat report.

\section{Motivation and Problem Statement}
\label{sec:motivation}

\subsection{Motivation}
Sigma rules are a widely adopted, vendor-agnostic signature format (written in YAML) for encoding CTI as detection logic across diverse SIEM platforms~\cite{suominen2024cyber}. This portability makes Sigma particularly valuable, but also makes relevant rule creation even more critical. In practice, security teams often need to write new Sigma rules to support proactive threat hunting and to address diverse operational contexts. Traditional methods for manually crafting Sigma rules are not only time-consuming and error-prone, but also require significant domain expertise, limiting their scalability and responsiveness to emerging threats. This manual workflow cannot scale and urgently calls for automation~\cite{alaeifar2024current}. Despite the existence of open-source Sigma rule repositories, these repositories cannot keep pace with the rapid evolution of adversarial tactics.

\begin{figure}[!h]
    \centering
    \includegraphics[width=0.9\linewidth]{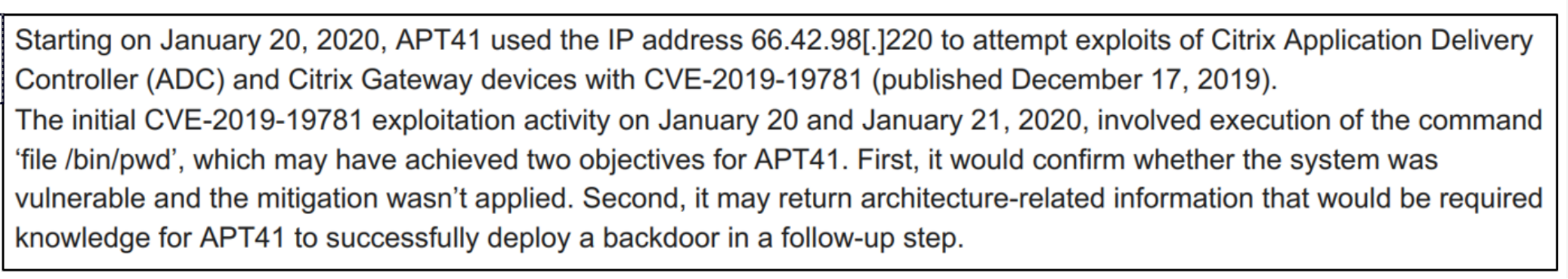}
    \caption{Excerpt from Mandiant's APT41 threat report~\cite{glyer2024apt41}. This example illustrates the unstructured nature and narrative style of CTI reports, which pose challenges for automated rule extraction.}
    \label{fig:report_example}
\end{figure}

CTI reports (\eg Fig.~\ref{fig:report_example}) are inherently free-form and heterogeneous, often blending detailed attack steps with benign observations and inconsistent formatting. Automated parsing of such reports into Sigma's structured conditions is extremely challenging. Prior research has proposed various NLP-driven extraction techniques, but these generally yield only limited structured outputs (such as identification of IoCs, or \mitre techniques) that still require extensive manual interpretation~\cite{husari2017ttpdrill, satvat2021extractor}. Moreover, even the existing Sigma rule corpus is far from complete: study~\cite{10.1145/3696410.3714798} found that only 60 of 193 \mitre patterns had a corresponding Sigma rule (approximately 31\% coverage). In other words, many behaviors documented in CTI may have no ready-made detection rule at all, forcing analysts to fill in gaps manually.

\begin{table*}[htbp]
\centering
\caption{{\sghaffar Comparison table of knowledge extraction solutions from CTI.}}
\label{tab:RelatedWork-comparison}
\resizebox{\linewidth}{!}{
\begin{tabular}{@{\hskip 0pt}lcccccc *{3}{c} *{4}{c}@{\hskip 0pt}}
\toprule
\textbf{Reference} & \textbf{Year} & \textbf{Technique} & \textbf{Dataset} & \textbf{Target Env.} & \textbf{Output Type} &
\multicolumn{3}{c}{\textbf{Output}} &
\multicolumn{2}{c}{\textbf{Augmentation}} &
\multicolumn{3}{c}{\textbf{Extraction}} \\
\cmidrule(lr){7-9} \cmidrule(lr){10-11} \cmidrule(lr){12-14}
& & & & & &
\textbf{Act.} & \textbf{Rob.} & \textbf{WF} &
\textbf{KBs} & \textbf{Temp.} &
\textbf{Rel.} & \textbf{IoCs} & \textbf{TTPs} \\
\midrule
\multicolumn{14}{c}{\textbf{NLP-based Solutions}} \\
\midrule
\textsc{TTPDrill}~\cite{husari2017ttpdrill} & 2017 & Unsupervised NLP & Symantec & On-premise & STIX & \ding{55} & \ding{55} & \ding{55} & \ding{55} & \ding{55} & \ding{51} & \ding{51} & \ding{51} \\
\textsc{ThreatRaptor}~\cite{gao2021enabling} & 2021 & Unsupervised NLP & DARPA TC & On-premise & Behavior Graph + SQL & \ding{55} & \ding{55} & \ding{55} & \ding{55} & \ding{55} & \ding{51} & \ding{51} & \ding{51} \\
\midrule
\multicolumn{14}{c}{\textbf{BERT-based and BiLSTM Solutions}} \\
\midrule
\textsc{CASIE}~\cite{satyapanich2020casie} & 2020 & BiLSTM & CyberWire & On-premise & Knowledge Graph & \ding{55} & \ding{55} & \ding{55} & \ding{55} & \ding{55} & \ding{51} & \ding{51} & \ding{55} \\
\textsc{Extractor}~\cite{satvat2021extractor} & 2021 & BERT-BiLSTM & Mixed Sources & On-premise & Behavior Graph & \ding{55} & \ding{55} & \ding{55} & \ding{55} & \ding{55} & \ding{51} & \ding{51} & \ding{51} \\
\textsc{Open-CyKG}~\cite{sarhan2021open} & 2021 & BiLSTM & MalwareDB & On-premise & Knowledge Graph & \ding{55} & \ding{55} & \ding{55} & \ding{55} & \ding{55} & \ding{51} & \ding{55} & \ding{51} \\
\textsc{SecIE}~\cite{park2022secie} & 2022 & BERT & CVE & On-premise & Knowledge Graph & \ding{55} & \ding{55} & \ding{55} & \ding{51} & \ding{55} & \ding{51} & \ding{51} & \ding{55} \\
\textsc{TriCTI}~\cite{liu2022tricti} & 2022 & BERT & Custom & On-premise & Labeled IoCs & \ding{55} & \ding{55} & \ding{55} & \ding{55} & \ding{55} & \ding{55} & \ding{51} & \ding{55} \\
\textsc{LADDER}~\cite{alam2024ladder} & 2023 & BERT & Custom & On-premise & Knowledge Graph & \ding{55} & \ding{55} & \ding{55} & \ding{51} & \ding{55} & \ding{51} & \ding{51} & \ding{51} \\
\textsc{CyberEntRel}~\cite{shah2024cyberentrel} & 2024 & RoBERTa-BiGRU-CRF & Custom & On-premise & Knowledge Graph & \ding{55} & \ding{55} & \ding{55} & \ding{51} & \ding{55} & \ding{51} & \ding{51} & \ding{55} \\
\textsc{FlowGuardian}~\cite{ahmadou2025automatedattacktestflowextraction} & 2025 & BERT & Custom & On-premise & TestFlow & \ding{55} & \ding{55} & \ding{55} & \ding{55} & \ding{55} & \ding{51} & \ding{51} & \ding{51} \\
\midrule
\multicolumn{14}{c}{\textbf{LLM-based Solutions}} \\
\midrule
\textsc{Purba}, \textsc{Chu}~\cite{purba2023extracting} & 2023 & GPT-3.5 & Twitter Posts & On-premise & Labeled IoCs & \ding{55} & \ding{55} & \ding{55} & \ding{55} & \ding{55} & \ding{55} & \ding{51} & \ding{55} \\
\textsc{acTion}~\cite{rastogi2023actionable} & 2023 & GPT-3.5 & Custom & On-premise & STIX & \ding{55} & \ding{55} & \ding{55} & \ding{51} & \ding{55} & \ding{51} & \ding{51} & \ding{51} \\
Liu~\etal~\cite{liu2023chatgpt} & 2023 & ChatGPT & Custom & On-premise & Knowledge Graph & \ding{55} & \ding{55} & \ding{55} & \ding{51} & \ding{55} & \ding{51} & \ding{51} & \ding{51} \\
\textsc{LLM-TIKG}~\cite{zhang2023few} & 2023 & LLaMA-2-7B (fine-tuned) & Custom & On-premise & Knowledge Graph & \ding{55} & \ding{55} & \ding{55} & \ding{51} & \ding{55} & \ding{51} & \ding{51} & \ding{51} \\
Fengrui~\etal~\cite{fengrui2024few} & 2024 & LLaMA-2-7B (fine-tuned) & ATT\&CK STIX Data & Hybrid & MITRE ATT\&CK TTPs & \ding{55} & \ding{55} & \ding{55} & \ding{51} & \ding{55} & \ding{51} & \ding{51} & \ding{51} \\
\textsc{IntelEX}~\cite{xu2024intelexllmdrivenattacklevelthreat} & 2024 & LLM & APT CTI & On-premise & Knowledge Graph + Sigma Rule & \ding{51} & \ding{55} & \ding{55} & \ding{51} & \ding{55} & \ding{51} & \ding{51} & \ding{51} \\
\textsc{LLMCloudHunter}~\cite{10.1145/3696410.3714798} & 2025 & GPT-4o & Custom & Cloud & Sigma Rules & \ding{51} & \ding{55} & \ding{55} & \ding{55} & \ding{55} & \ding{51} & \ding{51} & \ding{51} \\
\addlinespace[0.35em]
\midrule
\addlinespace[0.15em]
\textbf{\sol} & 2025 & LLM (LLM-as-a-Judge) & Custom & Cloud \& On-premise & Flow of Sigma Rules
& \ding{51} & \ding{51} & \ding{51} & \ding{51} & \ding{51} & \ding{51} & \ding{51} & \ding{51} \\
\addlinespace[0.2em]
\bottomrule
\end{tabular}
}
\vspace{-0.5em}
\begin{tabular}{l}
\textbf{Act.}: \textit{Actionable},
\textbf{Rob.}: \textit{Robust to input quality},
\textbf{WF}: \textit{Workflow},
\textbf{KBs}: \textit{External Knowledge Base},
\textbf{Temp.}: \textit{Template Retrieval},
\textbf{Rel.}: \textit{Relations}.
\end{tabular}
\end{table*}

\subsection{Problem Statement}
Most methods~\cite{10.1145/3696410.3714798, xu2024intelexllmdrivenattacklevelthreat, rastogi2023actionable} rely heavily on the quality and structure of the input report, which means they produce inconsistent results when faced with unstructured or poorly formatted documents. 

Other approaches~\cite{alam2024ladder, zhang2023few} lack contextual enrichment; they do not incorporate additional knowledge sources to augment the information in the report, therefore limiting the accuracy and relevance of the generated rules.

Moreover, the insights extracted by these systems often lack the granularity and structured format necessary for direct integration into SIEM environments. Consequently, security teams must manually refine or complete these partially generated rules, undermining the goal of full automation.

More fundamentally, there is a conspicuous gap in research when it comes to translating narrative threat intelligence into actionable content. Only a few studies~\cite{10.1145/3696410.3714798, xu2024intelexllmdrivenattacklevelthreat} have attempted to bridge this gap, and they face several obstacles:
\begin{enumerate*}[(i)]
    \item the inherent complexity of converting natural-language descriptions of attacks into the formal structure of Sigma rules,
    \item the difficulty of correlating the sequence of extracted attack steps with existing detection rule patterns, which is essential to ensure that any generated rule is consistent with known attack behaviors and truly relevant. 
\end{enumerate*}

These challenges have so far prevented the development of a solution capable of producing high-quality, ready-to-use Sigma rules from unstructured threat reports.

\section{Related Work}
\label{sec:relatedwork}
CTI plays a critical role in cybersecurity by providing timely insights into threat actors' tactics and techniques~\cite{alaeifar2024current}. 
In this section, we review existing solutions for CTI knowledge extraction and \textcolor{black}{Sigma} rule generation. 
Table~\ref{tab:RelatedWork-comparison} compares \sol and other tools for knowledge extraction from CTI reports.

\subhead{CTI Knowledge Extraction}
Early methods for knowledge extraction from CTI include \textsc{TTPDrill}~\cite{husari2017ttpdrill}, which uses unsupervised NLP to extract threat actions from Symantec reports and map them to \mitre techniques. \textsc{EXTRACTOR}~\cite{satvat2021extractor} advances this by using BERT-BiLSTM to convert unstructured CTI into formal evidence graphs for threat hunting. \textsc{ThreatRaptor}~\cite{gao2021enabling} combines action and indicator extraction into structured behavior graphs. \textsc{CASIE}~\cite{satyapanich2020casie} and \textsc{Open-CyKG}~\cite{sarhan2021open} use BiLSTM-based models to extract entities and relations from reports to populate threat knowledge graphs. These works structure threat knowledge but do not produce detection rules or offer mechanisms to assess correctness.
Recently, LLMs have been applied to improve knowledge extraction. \textsc{LADDER}~\cite{alam2024ladder} and \textsc{SHIELD}~\cite{gandhi2025shield} demonstrate entity extraction and threat reasoning using contextual modeling, with \textsc{SHIELD} employing iterative refinement to improve reliability.

\subhead{Sigma Rule Generation}
 
\textsc{LLMCloudHunter}~\cite{10.1145/3696410.3714798} uses GPT-4 to parse cloud-focused CTI reports and generate candidate Sigma rules. It shows high success in rule compilation and precision for cloud environments, but lacks a rule validation mechanism, making its outputs vulnerable to hallucinations. Furthermore, its design depends heavily on input quality and assumes well-parsed paragraphs, limiting robustness to noisy or complex CTI. Other systems, such as \textsc{SHIELD}~\cite{gandhi2025shield} and \textsc{acTion}~\cite{rastogi2023actionable}, apply LLMs for alert explanation or constructing technique knowledge graphs from structured threat sources. While relevant for enhancing threat visibility, they do not generate deployable Sigma detection rules. 

Similarly, recent works~\cite{purba2023extracting,zhang2023few,fengrui2024few,liu2023chatgpt} apply LLMs for IoC classification or knowledge base population but lack the full pipeline required for rule synthesis. \textsc{IntelEX}~\cite{xu2024intelexllmdrivenattacklevelthreat} is a notable system that leverages in-context prompting with external knowledge bases and a single-pass LLM-as-a-Judge mechanism to extract structured multi-step threat intelligence. It achieves strong performance on TTP and IoC extraction and shows early potential for generating Sigma rule candidates. However, \textsc{IntelEX} remains limited to content present in the report itself and does not perform template-based rule enrichment or similarity retrieval, which constrains rule generalizability and robustness.
 
In contrast, \sol introduces an end-to-end solution for CTI to Sigma translation. It performs contextual analysis and graph-based similarity matching to retrieve the most relevant Sigma rule templates, enriches them with external intelligence, and completes the rule using generative LLMs. A dual-LLM validation process (LLM-as-a-Judge) iteratively ensures syntactic correctness and semantic alignment with the CTI content. Unlike \textsc{LLMCloudHunter}, which performs one-pass generation on curated input, \sol is robust to formatting variations and enhances rule fidelity via knowledge-driven retrieval and multi-step verification. Unlike \textsc{IntelEX}, which remains constrained to within-report content, \sol goes beyond the report to build a complete and context-rich attack scenario grounded in known detection patterns.

\begin{figure*}[!ht]
    \centering
    \includegraphics[width=.9\linewidth]{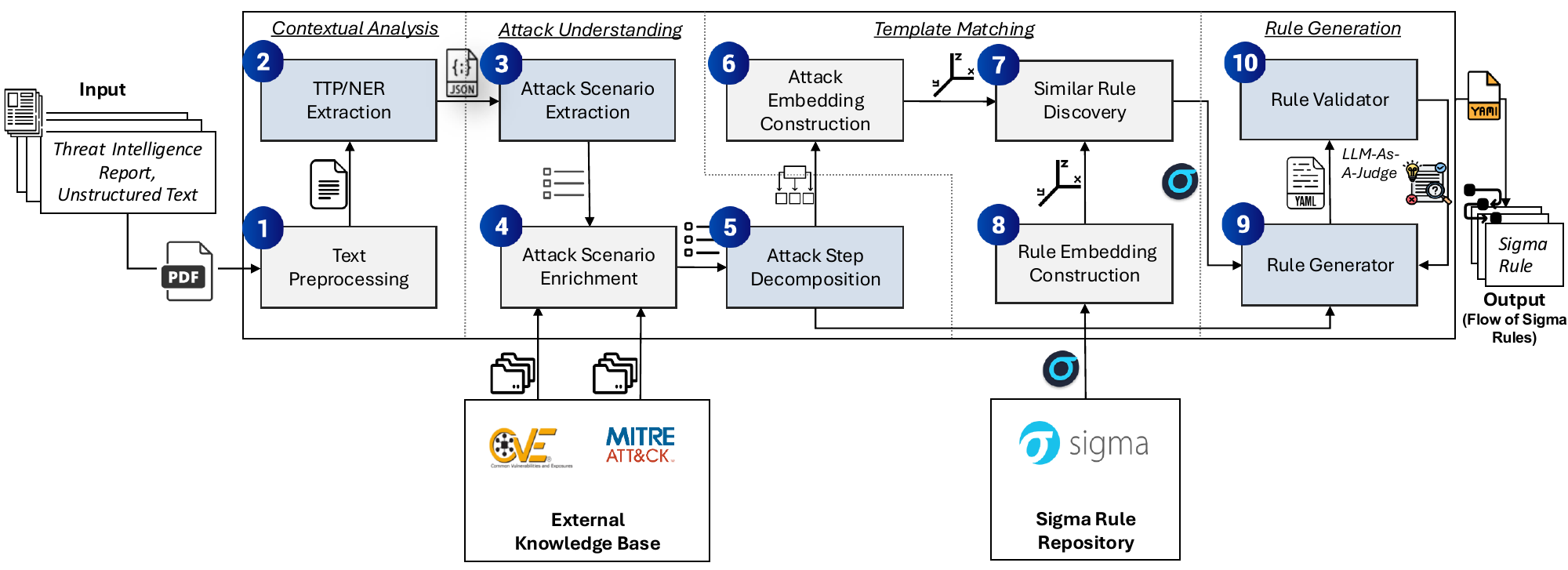}
    \caption{{\sghaffar High-level overview of \sol and its components (components shown by blue are based on LLM}).}
    \label{fig:AutoSigma_design}
\end{figure*}

\section{Threat Model and Assumptions}
\label{sec:threatmodel}
\subhead{In-scope threats} Our solution targets sophisticated, multi-stage cyberattacks, such as APTs, that are described in unstructured documents (\eg threat reports, incident response documentation). These attacks typically exploit software vulnerabilities (\eg Common Vulnerabilities and Exposures -- CVE), leverage known adversary tactics and techniques, and proceed through a kill-chain involving initial access, lateral movement, and impact.

\subhead{Out-scope threats} \sol does not aim to detect physical security incidents, insider threats not documented in reports, or attacks that do not leave observable artifacts suitable for Sigma rule generation. Similarly, zero-day threats with no previous contextual documentation are not within the immediate scope of detection unless indirectly referenced in CTI.

\subhead{System Assumptions}
We assume that the input is an unstructured threat report that describes one or more attack scenarios. Although our experiments focus on APT reports, our work is not limited to them; the input can be any textual description of malicious activity, including incident response reports, blog posts, or informal threat notes. Our methodology does not require the input text to be cleanly structured or annotated. We assume input reports are authentic and not intentionally manipulated.

We assume that each report may contain multiple discrete attack steps. As such, the system is expected to generate not one, but a sequence of Sigma rules, each corresponding to a specific attack action or tactic. These rules may vary in granularity depending on the level of detail available in the input. Finally, we assume the correctness of the external knowledge bases that the solution is leveraging.

\section{Automated Sigma Rules Generation}
\label{sec:systemarchitecture}

\subsection{System Overview}

\subsubsection{Overview:}
\sol, depicted in Fig.~\ref{fig:AutoSigma_design}, is designed to automate the transformation of APT reports into high-quality, actionable Sigma rules. The architecture follows a modular pipeline approach, consisting of four main components.

The first component of \sol is Contextual Analysis (Section~\ref{sec:contextual_analysis}), where it performs text preprocessing and applies Named Entity Recognition (NER) and TTP extraction to identify key elements within the unstructured threat report. Next, in the Attack Understanding component (Section~\ref{sec:attack_understanding}), the extracted entities and tactics are analyzed to infer possible attack scenarios. These scenarios are then enriched using external knowledge bases such as \mitre and NVD, allowing \sol to incorporate intelligence beyond what is explicitly stated in the report, thereby mitigating the limitations imposed by poor-quality or incomplete input. Subsequently, complex attacks are decomposed into atomic attack steps, enabling precise reconstruction of the attack chain and ensuring that generated rules are both comprehensive and granular.
Following this, the Template Matching component (Section~\ref{sec:template_matching}) employs a semantic similarity search over a curated repository of validated Sigma rules. By embedding both the extracted attack step and the rule corpus into a shared vector space, \sol retrieves the closest matching rule template, grounding its output in previously vetted detection logic. This retrieval-based grounding is a novel contribution that ensures generated rules inherit the structural correctness and semantic intent of battle-tested Sigma patterns, something not addressed in prior approaches.
Finally, in the Rule Generation component (Section~\ref{sec:rule_generation}), \sol uses an LLM-as-a-Judge paradigm to iteratively refine and validate each Sigma rule. A generation model proposes a candidate rule, while a validator model critiques and suggests corrections. This adversarial loop is repeated until the output satisfies both syntactic validity and semantic alignment with the source scenario. Through this pipeline, 
\sol combines multi-stage contextual reasoning, external enrichment, template-driven rule retrieval and iterative validation to produce robust Sigma rules that are both immediately deployable and semantically faithful to the underlying threat report.

{\sghaffar
\subsubsection{Novelty:}
\sol introduces multiple novel elements that distinguish it from prior approaches. First, it performs end-to-end automation of the Sigma rule generation process, starting from raw unstructured threat reports to producing relevant, deployable rules. Second, it integrates external knowledge sources such as the \mitre knowledge base and the National Vulnerability Database (NVD){\footnote{NVD: \url{https://nvd.nist.gov/}}} to enrich attack context and fill in gaps from incomplete reports. Third, it constructs the entire attack chain by decomposing each scenario into a sequence of attack steps, each mapped to a Sigma rule. Fourth, it leverages an up-to-date repository of official Sigma rules to retrieve and adapt existing templates through similarity search. Finally, a dual-LLM feedback loop, where a rule generator and a rule validator iteratively collaborate to refine the quality of the generated \textcolor{black}{Sigma} rules.

\subsubsection{Design Challenges:}
Designing \sol required addressing several challenges. CTI reports are typically written in \textcolor{black}{free form} and lack consistent structure, making it difficult to directly extract entities or behavior sequences. The availability of high-quality, labeled cybersecurity data is limited, which constrains the use of traditional supervised models for tasks like NER. Additionally, large language models, while powerful, are prone to hallucinations or producing outputs that are syntactically incorrect or semantically misaligned with the input. Controlling these errors and handling failure cases gracefully was essential to ensuring pipeline reliability. \sol addresses these challenges through contextual analysis, leveraging external knowledge bases, \textcolor{black}{and} using LLMs for tasks such as NER extraction\textcolor{black}{. Additionally, the} LLM-as-a-Judge refinement loop improves the quality of the final results.
}

\subsection{Contextual Analysis}
\label{sec:contextual_analysis}

\begin{definition}
\label{def:contextual_analysis}
\sepehr{Let $\mathcal{D}$ represent the set of unstructured CTI documents. We define $\mathcal{S}$ as the set of preprocessed text segments (derived via Component \navycircle{1} in Fig.~\ref{fig:AutoSigma_design}) and $\mathcal{X}$ as the multidimensional set of extracted cybersecurity entities, including NERs, IoCs, and \mitre TTPs (derived via Component \navycircle{2}). The resulting contextual representation space is defined as $\mathcal{C} \subseteq 2^{\mathcal{S}} \times 2^{\mathcal{X}}$. The contextual analysis stage is modeled as a mapping function $F_{\text{ctx}}: \mathcal{D} \to \mathcal{C}$, such that for any report $d \in \mathcal{D}$, the function yields a structured tuple $c_d = (S_d, X_d)$, where $S_d$ captures the preprocessed narrative and $X_d$ isolates the contextual information.}
\end{definition}

The first component of \sol is responsible for extracting meaningful threat intelligence from {\sghaffar raw threat reports}. Since these reports vary widely in structure and format, this module applies preprocessing techniques to ensure consistency, remove irrelevant content, and isolate the information most pertinent to subsequent analysis.

\subsubsection{Text Preprocessing:}

\sol ingests threat reports and performs rigorous text preprocessing, which prepares the data for further analysis. This phase involves several NLP techniques~\cite{10128641} to refine raw, unstructured data: 

\begin{enumerate*}[(a)]
    \item extracting the text given a threat report;
    \item text cleaning, which removes the irrelevant data from the text and prepares the text for further analysis; 
    \item normalization, which standardizes the data for consistency and improved algorithm performance; and 
    \item segmentation, which allows us to feed the language models with input in manageable pieces, optimizing the performance of the LLMs in our architecture by providing context segment-by-segment instead of in one large block~\cite{hope_chunking}.
\end{enumerate*}

\subsubsection{TTP and NER Extraction:}
\label{sec:ner_ttp}
Various NLP solutions have been proposed for extracting named entities and mapping tactics, techniques in cybersecurity contexts, such as PELAT~\cite{lin2022attack}, SMET~\cite{basel2023smet}, and Adema~\cite{li2024automated}. However, these approaches often rely on annotated datasets and may lack adaptability to diverse, unstructured threat reports. To address these limitations, \sol uses LLM for NER extraction, leveraging its capabilities to generalize from limited data and understand complex contexts. We also experimented with a BERT-based NER model~\cite{devlin2019bertpretrainingdeepbidirectional}, but the LLM approach yielded better results given the limited training data available for these entity types. Once the key entities are extracted from the report, \sol analyzes them to find tactics, techniques, and aligns the information with the \mitre knowledge base. This alignment provides a precise categorization of the described attacker behaviors, ensuring that subsequent processing steps have a structured understanding of the threat context.

\begin{example}
Consider the following excerpt from a threat report:

\textit{"APT41 leveraged CVE-2019-19781 to deploy Cobalt Strike beacons, facilitating lateral movement across the network."} 

In this case, \sol identifies APT41 as the threat actor, CVE-2019-19781 as the exploited vulnerability, and Cobalt Strike as the malware used in the operation. Based on this, it infers the corresponding \mitre techniques: Exploit Public-Facing Application (T1190), Command and Control through Beaconing (T1071.001), and Lateral Movement via Remote Services (T1021).
\end{example}

\subsection{Attack Understanding}
\label{sec:attack_understanding}

\begin{definition}
\label{def:attack_understanding}
\sepehr{Let $\mathcal{A}$ represent the set of enriched and decomposed attack steps derived from Components \navycircle{3}, \navycircle{4}, and \navycircle{5}. Each element $a \in \mathcal{A}$ describes a single malicious behavior together with its associated indicators, \mitre annotations, and enriched information (\eg CVE metadata). The attack understanding phase is modeled as a transformation $f_{\text{au}} : \mathcal{C} \to 2^{\mathcal{A}}$, which maps a contextual representation $c_d \in \mathcal{C}$ to a discrete, ordered set of enriched and decomposed attack steps $A_d = \{a_1, a_2, \dots, a_n\} \subseteq \mathcal{A}$.}
\end{definition}

{\sghaffar While NER extraction and \mitre technique mapping provide a foundational understanding of threat reports, they are often insufficient for generating high-quality, specific Sigma rules. This is because the raw extracted entities alone may lack context, depth, or completeness, especially in cases where the report is vague or incomplete. Therefore, \sol performs an additional layer of analysis to construct a more coherent and actionable representation of the attack.
In this phase, \sol goes beyond simple extraction, meaning it not only identifies entities and techniques, but also enriches them using external cybersecurity sources. For example, if a CVE is mentioned, \sol queries the National Vulnerability Database (NVD)\footnote{NVD: \url{https://nvd.nist.gov/}} to retrieve its detailed description, severity, and affected software. Similarly, if a known threat actor is referenced, information from the \mitre knowledge base is used to retrieve related tactics, tools, and historical campaigns. Furthermore, to ensure that the Sigma rules generated are both granular and actionable, this component decomposes complex attack descriptions into a sequence of atomic \textcolor{black}{attack steps}. Each \textcolor{black}{step} corresponds to a specific malicious behavior or tactic, and ultimately maps to a separate Sigma rule. This decomposition is critical for generating accurate and modular detection logic.}

\subsubsection{Attack Scenario Extraction:}
Using the identified NER entities and TTPs, the system reconstructs meaningful attack scenarios mentioned in the report. This step synthesizes fragmented information into structured attack scenarios.
For example, APT41 exploited CVE-2019-19781 to gain initial access to the target system by utilizing FTP to download a malicious payload.

These attack scenarios provide structured context for further enrichment and Sigma rule generation.
By the end of this step, \sol produces a structured representation of the extracted intelligence, making it suitable for subsequent enrichment and detection rule generation.

\subsubsection{Attack Enrichment:}
To provide additional context to attack scenarios, \sol queries multiple cybersecurity knowledge bases. The enrichment process involves, but \textcolor{black}{is} not limited to\textcolor{black}{,} the following:

\begin{itemize}
\item \textit{CVE Enrichment:} If an attack uses a CVE (\eg CVE-2019-19781), \sol retrieves relevant details such as the description of the vulnerability, CVSS score and severity rating, affected software, and versions. {\sghaffar This information is sourced from external knowledge bases. In this work, we used the CVE repository as our knowledge base. The repository contains over 277,000 CVE records, each providing standardized information about publicly known cybersecurity vulnerabilities, including descriptions, severity scores, and affected products~\cite{cve_org}.}

\item \textit{Threat Actor Enrichment:}

{\sghaffar We also use \mitre as a knowledge base, where \sol queries it to extract additional information related to the attack. This information can be captured in different components, such as tools, malware, and software that a threat actor might use.}
\end{itemize}

\subsubsection{Attack Step Decomposition:}
\label{sec:attack_decomposition}
{\sghaffar Given that an attack scenario often contains multiple attack behaviors or subgoals, it is not feasible to generate a single Sigma rule that effectively captures all of them. Therefore, \sol decomposes each enriched scenario into smaller, discrete attack steps. {\sghaffar This decomposition plays a critical role in aligning each step with existing detection logic more effectively. Unlike prior approaches~\cite{10.1145/3696410.3714798, xu2024intelexllmdrivenattacklevelthreat} that attempt to encapsulate an entire attack scenario within a single rule, our method breaks down complex threats into distinct, manageable components, enabling the generation of multiple, focused flows of \textcolor{black}{Sigma} rules. This not only enhances the precision of each rule but also promotes modularity and scalability in detection coverage.
} Each step isolates a specific tactic or action taken by the adversary, which in the next phase will be mapped to a detection rule. 
To perform this decomposition, \sol uses a prompting-based approach with an LLM, where the enriched attack description is input and the model is guided to split the scenario into atomic substeps. This step ensures that every component of an attack is isolated and can be individually analyzed for detection logic, thereby eliminating ambiguity and improving the fidelity of the final detection rules. An example of the output for this step is illustrated in Appendix~\ref{sec:output_example}. \textcolor{black}{The} prompting strategy is \textcolor{black}{also} discussed in Appendix~\ref{sec:prompts}.}

\subsection{Template Matching}
\label{sec:template_matching}

\begin{definition}
\label{def:template_matching}
    \sepehr{Let $\mathcal{R}$ denote the set of all Sigma rules within the SigmaHQ repository. The template matching phase is defined as a retrieval function $f_{\text{tm}} : \mathcal{A} \to 2^{\mathcal{R}}$. For any enriched attack step $a \in \mathcal{A}$, the function yields a non-empty, finite subset $T_a \subseteq \mathcal{R}$, representing the candidate templates that exhibit the highest semantic similarity to $a$ within the shared vector space $\mathcal{V}$ (derived via Components \navycircle{6}, \navycircle{7}, and \navycircle{8}). This is expressed as $T_a = \{r \in \mathcal{R} \mid \text{sim}(a, r) > \tau\}$, where $\text{sim}$ is the BERTScore metric and $\tau$ is a relevance threshold.}
\end{definition}

{\sghaffar To improve the relevance and quality of the generated rules, we introduce a novel step: retrieving similar existing Sigma rules to be used as templates for generating new ones.} In the third stage, \sol prepares to leverage existing knowledge of known detection rules by embedding both the attack steps derived from the report and a large collection of known and valid Sigma rules into a common vector space. By representing attack descriptions extracted from the attack understanding phase, and Sigma rules as numerical embeddings, \sol can perform efficient similarity searches to identify which known rules are most relevant to the attack steps at hand. This component bridges the gap between the enriched attack intelligence and the rule generation phase by providing a starting template for each detection rule \sol needs to generate.

\subsubsection{Attack Embedding Construction:}
Each attack description produced in the previous stage is converted into a vector embedding using a BERT-based model~\cite{devlin2019bertpretrainingdeepbidirectional}. This transformation captures the semantic meaning of the attack description in a high-dimensional space. The motivation behind this step is to enable comparison with existing Sigma rule embeddings, by representing both the attack descriptions and Sigma rules in the same vector space.

\subsubsection{Rule Embedding Construction:}
\label{sec:sigma_repo_representation}

\sol uses an up-to-date collection of Sigma rules, drawn from the latest SigmaHQ\footnote{SigmaHQ: \url{https://github.com/SigmaHQ/sigma}} repository. To organize this repository for efficient search, we built a graph-based representation of those Sigma rules. \sepehrgh{This representation acts as an indexing method that accelerates rule retrieval during the next stage of \sol. In SigmaHQ repository, rules are already arranged hierarchically into categories such as \textit{application}, \textit{cloud}, \textit{linux}, and \textit{macos}, each containing multiple subcategories including \textit{linux} $\rightarrow$ \textit{auditd}, \textit{file\_event}, \textit{process\_creation} and \textit{cloud} $\rightarrow$ \textit{aws}, \textit{azure}, \textit{gcp}, \textit{m365}. We convert this directory structure into a lightweight graph in which top-level categories form the first layer, subcategories form the second layer, and individual Sigma rules form the third layer. This abstraction enables efficient traversal and indexing of the rule set during the next stage.} \sepehr{
We formally define this structure as a directed acyclic graph (DAG) G = (V, E), where nodes represent the repository root, platforms, categories, and individual Sigma rules, and edges encode hierarchical relationships between them. The root node corresponds to the SigmaHQ repository, followed by platform nodes (e.g., linux, windows, cloud), category nodes (e.g., process\_creation, file\_event), and finally rule nodes.
During retrieval, given an attack step, \sol first narrows the search space by selecting relevant platform nodes, then traverses to their corresponding categories, and finally evaluates candidate rules within those categories using semantic similarity. This hierarchical traversal reduces the search space and improves retrieval efficiency.
}
Because the SigmaHQ repository is continuously updated, \sol automatically fetches the most recent set of Sigma rules whenever updates are available. \sepehr{With each execution of the solution, \sol refreshes the SigmaHQ repository to ensure that the most updated rules are incorporated into the pipeline.} It then incorporates these into its knowledge base, ensuring that even cutting-edge detection techniques are considered during rule generation. Deprecated rules are pruned and new rules are indexed without manual intervention, enabling \sol to build on top of the recent rule set. We leverage this repository to find existing Sigma rules that closely match the attack step in question. If a relevant rule is found, we refine and build upon it. Otherwise, we use the rules as one-shot learning examples for the LLM.

\subsubsection{Similar Rule Discovery:} The repository of Sigma rules serves as the foundation for the similarity matching process. With every rule indexed and embedded, \sol can perform rapid searches to find which existing rules are most relevant to a given \textcolor{black}{attack step}. Moreover, because the repository is kept current, the framework ensures that it is comparing against the latest known detection patterns. This means the output rules will align closely with community-curated detection techniques and will not reinvent rules that already exist in the repository. In order to achieve this, we compute BERTScore~\cite{zhang2020bertscoreevaluatingtextgeneration} between an attack step embedding and Sigma rule embeddings to measure how closely a given attack description matches the scope of any known detection rule. {\sghaffar The reason why we use BERTScore instead of other similarity scores, such as cosine similarity or BLEU~\cite{10.3115/1073083.1073135}, is that this method allows us to capture not only exact word matches but also the overall meaning, fluency, and order of the output, making it more effective than traditional n-gram-based metrics like BLEU.} This embedding-based approach allows the system to efficiently scan a large corpus of rules and pinpoint those that are contextually most similar to the new attack scenario.

\subsection{Rule Generation}
\label{sec:rule_generation}

\begin{definition}
\label{def:rule_generation}
\sepehr{Let $\sigma$ denote the set of syntactically valid Sigma rules. The rule generation component (Component \navycircle{9}) and its iterative validation loop (Component \navycircle{10}) are modeled as a refinement function $f_{\text{rg}}: \{(a, T_a) \mid a \in \mathcal{A}, T_a \subseteq \mathcal{R}\} \to 2^{\sigma}$. For a given enriched attack step $a$ and its associated template set $T_a$, the function outputs a set of finalized rules $R_a \subseteq \sigma$. A rule $r$ is included in $R_a$ only if it satisfies the predicate $\text{valid}(r) \land \text{aligned}(r, a)$, where these conditions are validated through the LLM-as-a-Judge refinement loop.}
\end{definition}
In the final phase, \sol takes the candidate Sigma rule templates identified in the previous stage and refines them into high-fidelity detection rules tailored to the specific attack scenarios.

\subsubsection{Rule Generator:} 
\label{sec:rule_generator}
The rule generation component is designed to transform enriched \textcolor{black}{attack steps} into high-quality, executable Sigma rules.
The primary goal of this stage is to take the best-matching Sigma templates (see \textcolor{black}{S}ection~\ref{sec:template_matching}) and enrich them for the given attack description while ensuring the resulting rule meets several key criteria. It must accurately reflect the attack techniques described in the source reports and cover all relevant IoCs along with the associated adversary behaviors. At the same time, the rule should maintain strong detection capabilities with minimal false positives, and it must strictly adhere to the Sigma rule format for seamless integration with SIEM platforms. To fulfill these objectives, \sol integrates \textcolor{black}{an} LLM-as-a-Judge mechanism~\cite{gu2025surveyllmasajudge} in which two LLMs work in an adversarial collaboration to iteratively refine the Sigma rules being generated. {\sghaffar Details are provided in Appendix~\ref{sec:prompts}.}

\subsubsection{LLM-as-a-Judge:} The rule generation phase consists of two distinct LLMs, {\sghaffar they can be the same underlying LLMs or they can be two different LLMs}, interacting with each other:

\subhead{1. Rule Generator (LLM-1)} 
This first LLM (the ``rule generator'') takes the retrieved Sigma rule template along with the context of the attack step and generates an initial Sigma rule tailored to the scenario. Essentially, LLM-1 fills in or modifies the template's fields (such as the rule title, description, detection section filters, and tags) using the information from the attack step. The output at this stage is a draft rule hypothesized to detect the described attack. 

\subhead{2. Rule Validator (LLM-2)}
The second LLM acts as a validator or \emph{judge}. It examines the proposed Sigma rule from LLM-1 and evaluates it against a set of predefined quality checks: 

\begin{itemize}[left=1pt]
    \item Does the rule accurately detect the described attack technique and behaviors? 
    \item Is the detection logic valid, sensible, and free of obvious gaps or mistakes (\eg correct field names, appropriate use of conditions)? 
    \item Is the rule properly formatted according to YAML and Sigma specifications?
\end{itemize} 
Based on this evaluation, LLM-2 provides a quantitative score or verdict on the rule's quality and generates structured feedback highlighting what needs improvement. For example, the validator might point out that a certain condition is too broad (risking false positives) or that a required field is missing or misformatted. 
\textcolor{black}{The interplay of these two models, one generating and the other validating, forms an adversarial learning loop.} LLM-1 and LLM-2 do not operate in isolation but rather communicate through the intermediate Sigma rule drafts and feedback. 

Iterative Refinement Process: Using the generator-validator setup above, \sol iteratively refines each Sigma rule through multiple rounds, as follows: 
\begin{enumerate}[left=1pt]
    \item \textbf{Initial Draft:} LLM-1 generates an initial enriched rule based on the attack description and the selected template (as described in \textcolor{black}{S}ection~\ref{sec:rule_generator}). 
    
    \item \textbf{Validation and Scoring:} LLM-2 evaluates this draft, checking its accuracy, logic, and format, and returns a score along with feedback on how to improve the rule. 
    
    \item \textbf{Feedback-Driven Revision:} LLM-1 incorporates LLM-2's feedback and revises the Sigma rule, for example, by tightening detection conditions, adding missing IoCs, or correcting format errors. 
    
    \item \textbf{Re-Validation:} The updated rule is sent back to LLM-2 for another round of evaluation. 
    
    \item \textbf{Repeat Until Optimal:} Steps 2-4 repeat, with the rule steadily improving each time, until the validator LLM judges that the rule meets the desired quality threshold. In practice, this loop converges after a few iterations, resulting in a rule that LLM-2 can no longer fault on the predefined criteria. 
\end{enumerate} 
This refinement loop continues until the Sigma rule is both syntactically valid and semantically robust for detection. \sepehr{One important note is that the LLM-as-a-Judge mechanism is not designed to handle hallucinations of the LLMs; rather, its purpose is to improve the quality of the generated output. Hallucination control is instead addressed through the use of lower temperature settings and fine-tuned LLMs (see Section \ref{sec:res}, RQ3.3). Additionally, we observed in very few cases (1-2 instances) that the refinement loop could get stuck in an infinite loop. To handle this, we introduced a maximum iteration limit of 5 in the pipeline.}

\section{Implementation}
\label{sec:implementation}
\sepehrgh{We build a Proof of Concept (PoC)}\footnote{\sepehrgh{\sol's Demo: \url{https://youtu.be/iSr6IurQ6BM}}} using Python 3 programming language. Our experiments were conducted on Intel Xeon E312xx, 6 vCPU @ 2.69Ghz, with 32GB RAM running Ubuntu 20.04. 
Further details regarding the PoC are provided in Appendix~\ref{sec:PoC}.

\subsection{LLM Models}
There are generally two methods to work with LLMs: (i) using a cloud-based LLM, trained and deployed by a third-party, or (ii) using a locally hosted LLM. Each approach has its own advantages and disadvantages. 

Cloud-based LLM access does not require extensive computational resources and simplifies request handling; however, it is often associated with usage costs, and concerns remain about data confidentiality and the possible influence of concurrent API requests on output quality. Running LLMs locally ensures data privacy and eliminates risks associated with multi-tenant environments, but demands significant computing resources. Without sufficient hardware, one must use lighter LLM variants, which may negatively impact performance.
For our evaluation, we used both methods:
\begin{itemize}[left=1pt]
    \item \textit{Local-based LLM:} We employed \textit{Lily-Cybersecurity-7B-v0.2}\footnote{Lily Cybersecurity Model: \url{https://huggingface.co/segolilylabs/Lily-Cybersecurity-7B-v0.2}}, a fine-tuned variant of Mistral-7B-Instruct-v0.2, specifically trained for cybersecurity tasks. 
    The training encompassed a broad spectrum of cybersecurity domains, including but not limited to APT management, malware analysis, incident response, and secure software development lifecycle. We configured the model with a temperature of 0.1 to minimize randomness and a maximum token length of 512 to ensure concise and relevant outputs.
    \item \textit{Cloud-based LLM:} We used \textit{ChatGPT-4o-mini}\footnote{ChatGPT 4o-mini: \url{https://platform.openai.com/docs/models/gpt-4o-mini}} \sepehrgh{and \textit{Llama-3.3-70b}\footnote{Llama-3.3-70B: \url{https://huggingface.co/meta-llama/Llama-3.3-70B-Instruct}} through OpenAI and Llama's APIs.}
    We also set the temperature parameter to 0.1 and the maximum token length to 512 to maintain consistency in output quality and length.
\end{itemize}

\sepehrgh{More details about the LLMs \textcolor{black}{are} provided in Appendix~\ref{sec:llm_specification}.}

\subsection{Datasets}
\label{sec:dataset}
\subhead{Threat Reports}
We use threat reports from the APTNotes repository\footnote{APTNotes repository: \url{https://github.com/aptnotes/data/}}, which gathers different threat intelligence from different vendors (\eg Mandiant, CrowdStrike). 
Given the lack of a comprehensive dataset, 
we focused only on APT41, APT28, and APT29, yet the solution can work with any report. 
Appendix~\ref{sec:apt41_summary} provides a summary of the CTI reports used in our evaluation.
\sepehrgh{To further evaluate different solutions, we also used the dataset released by \textsc{LLMCloudHunter}~\cite{10.1145/3696410.3714798}, comprising 20 cloud-security blogs (Azure/AWS/GCP) paired with their manually crafted Sigma rules. }

\subhead{Knowledge Base}
\sol incorporates two external knowledge bases. The NVD, which provides a comprehensive repository of CVEs. We utilize this database to extract relevant information about identified vulnerabilities. 
The second is the official SigmaHQ repository, which \sepehrgh{contains more than 3{,}600 rules spanning over 60 detection domains. These rules are distributed across several major categories, including but not limited to, operating system domains (\eg Windows, Linux, macOS), cloud environments (\eg AWS, Azure, GCP, M365) and more. This diversity reflects the breadth of the Sigma rule ecosystem.}

\section{\sepehrgh{Evaluation}}
\label{sec:res}

\sepehrgh{
We adopt three research questions that jointly assess \sol quantitatively and qualitatively: (i) benchmarking against state-of-the-art solution; (ii) cross-APT evaluation against LLM models, and (iii) design-choice and robustness studies isolating the effects of the LLM-as-a-Judge loop and model variation and sampling temperature.}

\begin{RQ*}{1}
How effective is \sol in generating high-quality and behavior-aligned Sigma rules compared to a state-of-the-art solution?
\end{RQ*}

\sepehrgh{We evaluate \sol, using \textsc{LLMCloudHunter} dataset, in two configurations \sol using cloud-based LLMs and \sol using a Local LLM, and compare against \textsc{LLMCloudHunter}.}

\sepehrgh{All systems are assessed with seven complementary metrics:
\begin{enumerate*}[label=(\roman*)]
    \item \textit{No. Rules:} number of generated Sigma rules per report;
    \item \textit{Rule Validity:} percentage of rules that are syntactically valid and convertible to SIEM queries (\eg Splunk/ELK) using sigma-cli\footnote{SigmaCLI: \url{https://github.com/SigmaHQ/sigma-cli}};
    \item \textit{Condition Accuracy:} focuses on the correctness of the condition fields, which specify the relationship between various selection fields.
    \item \textit{Rule Relevancy:} semantic similarity between each report and its generated rule set via BERTScore~\cite{zhang2019bertscore}.
    \item \textit{Partial Coverage:} number of generated rules that match at least one of the same \mitre techniques as a ground truth rule;
    \item \textit{Full Coverage:} number of generated rules whose detection logic is equivalent to a ground truth rule;
    \item \textit{Newly Discovered:} number of new rules generated by the model but not present in the ground truth;
\end{enumerate*}}
\sepehrgh{The results are summarized in Table~\ref{tab:benchmark_results} and discussed in the following sub-research questions.}

\begin{table}[h]
\centering
\caption{\sepehrgh{Average results of \sol against \textsc{LLMCloudHunter} using cloud-security blogs.}}
\label{tab:benchmark_results}
\renewcommand{\arraystretch}{1.1}

\begin{tabular}{lccc}
\toprule
& \makecell{\sol \\ (Cloud-based\\ OpenAI)}
& \makecell{\sol \\ (Local-based\\ Lily)}
& \makecell{LLMCloudHunter} \\
\midrule
No. Rules           & 21    & 25     & 8   \\
Rule Validity       & 100   & 77.94  & 94  \\
Condition Accuracy  & 100   & 100  & 100  \\
Rule Relevancy      & 80.4  & 77.7   & 76.4 \\
Partially Covered   & 97.40 & 84.41  & 80.9 \\
Fully Covered       & 83    & 74.93  & 74   \\
Newly Discovered    & 12    & 11     & 2    \\
\bottomrule
\end{tabular}
\end{table}

\sepehrgh{\textbf{RQ1.1. Are the generated rules syntactically correct, semantically consistent with the input reports?}}
\sepehrgh{Results for total number of rules and rule validity are shown in Table~\ref{tab:benchmark_results}. As we can see, \sol in both cloud-based and local-based, outperformed the state-of-the-art solution. Due to the fact that \textsc{LLMCloudHunter} tends to generate the rules based on the telemetry information, such as IoCs mentioned in the report, this limits the overall number of rules produced.
For rule validity, the cloud-based version of \sol achieves a score of 100\%, meaning every generated rule is syntactically valid and can be directly converted into a SIEM query. 
The local version, while producing a slightly lower validity percentage compared to \textsc{LLMCloudHunter}, primarily because it uses a lighter-weight LLM, still maintains a higher count of valid rules than \textsc{LLMCloudHunter} due to its larger output volume. 
This difference between the cloud-based and local-based configurations stems from their underlying LLM capabilities and generation precision, as further discussed in Appendix~\ref{sec:local_vs_cloud}.
Furthermore, the condition accuracy metric shows that all generated rules, across all evaluated models, achieve 100\% correctness in their condition fields. This indicates that the logical structure of each rule faithfully captures the intended detection behavior and can correctly operationalize the threat activity described in the reports.}

\sepehrgh{\textbf{RQ1.2. Do the generated rules correspond to behaviors explicitly described in the original reports?}}
\sepehrgh{In this experiment, all models generate rules that remain semantically aligned with their corresponding CTI reports. 
However, as the rule relevancy (Table~\ref{tab:benchmark_results}) shows that both cloud-based and local-based versions of \sol, achieve higher semantic similarity scores (80.4\% and 77.7\%, respectively) than the state-of-the-art solution (76.4\%). 
This demonstrates \sol's strength in maintaining contextual coherence throughout its Attack Scenario Extraction and Enrichment. 
Even with a higher number of generated rules, which typically increases the risk of semantic drift, \sol preserves alignment with the original text due to its contextual analysis and LLM-as-a-\textcolor{black}{J}udge components, ensuring the final outputs remain faithful to the report’s intent.
}

\sepehrgh{\textbf{RQ1.3. To what extent do the rules generated capture the threat behaviors and \mitre techniques documented in the reports?}}
\sepehrgh{As shown in Table~\ref{tab:benchmark_results}, both versions of \sol stand ahead of \textsc{LLMCloudHunter}. In terms of partially covered rules, \sol has covered 97.4\% and 84.4\%, versus 80.9\% for \textsc{LLMCloudHunter}. For fully covered rules, \sol has 83.0\% compared to 74.0\% obtained by \textsc{LLMCloudHunter}. 
This is due to the fact that \textsc{LLMCloudHunter} tends to focus on telemetry information such as IoCs when generating Sigma rules, which limits the overall number of rules produced. 
Moreover, \textsc{LLMCloudHunter} is specifically designed to filter cloud-based threat scenarios and therefore only generates cloud-based rules. 
Overall, all models perform reasonably well since they rely on the \mitre technique extraction at some stage of their process. 
However, \sol demonstrates superior coverage by 8\%, because its design does not have any domain constraint filters. Unlike \textsc{LLMCloudHunter}, \sol identifies techniques across diverse contexts, resulting in broader and more comprehensive coverage.
}

\sepehrgh{\textbf{RQ1.4. Can we identify new or previously uncovered Sigma rules?}}
\sepehrgh{The newly discovered metric highlights the capacity of a model to uncover attack behaviors not explicitly reflected in the ground truth. 
Manual rule creation often overlooks such subtle variations or decomposed steps. 
\sepehr{Importantly, these newly discovered rules are not arbitrary, but correspond to attack behaviors described in the CTI reports that are not captured by other baselines. As such, they contribute additional detection coverage by identifying relevant attacks that would otherwise remain undetected.} As shown in Table~\ref{tab:benchmark_results}, \sol achieves an average newly discovered rule of 12 compared to \textsc{LLMCloudHunter} of 2, as \sol employs attack scenario extraction and decomposition components, which break complex attack narratives into simpler attacks that are more effectively expressed as Sigma rules. \sol not only focuses on telemetry information described in the threat report, but also leverages external knowledge bases.
This process enables \sol to discover a greater number of novel rules (12 and 11 on average) compared to \textsc{LLMCloudHunter}.
}

\begin{RQ*}{2}
How does \sol perform relative to LLM baseline models in generating Sigma rules for APT reports?
\end{RQ*}

\sepehrgh{
To evaluate the performance of \sol against LLM models, we defined four metrics:
\begin{enumerate*}[label=(\roman*)]
    \item \textit{Coverage:} evaluate the content retention from the report at two levels:
    \begin{enumerate*}[(a)]
        \item \emph{IoC coverage} by checking if IoCs mentioned in the report have been covered by the generated rules, and
        \item \emph{\textsc{MITRE} coverage} by checking the covered techniques from the report and the generated rules. For this experiment we used Threat2Mitre~\cite{Threat2Mitre2024} as our baseline;
    \end{enumerate*}
    \item \textit{Relevancy:} Semantic alignment between each report and its generated rule set using BERTScore embeddings;
    \item \textit{Validity:} rules that are syntactically valid and convertible to SIEM queries (\eg Splunk/ELK) using sigma-cli; and
    \item \textit{Adaptability:} Pairwise BERTScore similarities among \emph{inputs} (reports) and among \emph{outputs} (rules).
\end{enumerate*}
}
\sepehrgh{It is important to mention that \textsc{LLMCloudHunter} is designed for cloud-specific content and does not produce any rules if the content is not related to cloud. Thus, we excluded it in this evaluation since our focus is on APT reports.}

\sepehrgh{\textbf{RQ2.1. Does \sol achieve higher coverage of described behaviors?}}
\sepehrgh{Table~\ref{tab:sigmarules_detailed_comparison} shows the number of Sigma rules generated per report. On average, ChatGPT-off-the-shelf \sepehr{(version: gpt-4o\footnote{gpt-4o: \url{https://developers.openai.com/api/docs/models/gpt-4o}})} generates only 6 rules per report, which is often insufficient given the sophistication and multi-stage nature of APT attacks. In contrast, \sol, under different configurations, because its system design, specifically its attack scenario decomposition component, consistently produces a higher number of rules, around 28 for local models and 24 for Cloud-based models.
}

\begin{table}[!h]
\centering
\caption{Number of Sigma rules generated per APT41 reports.}
\label{tab:sigmarules_detailed_comparison}
\renewcommand{\arraystretch}{1.12}
\setlength{\tabcolsep}{2pt}
\resizebox{\columnwidth}{!}{
\begin{tabular}{l c cc c}
\toprule
\multirow{2}{*}{Report}
& \multicolumn{2}{c}{Cloud-Based LLMs}
& \multicolumn{1}{c}{Local-Based LLMs}
& \multicolumn{1}{c}{Baseline} \\
\cmidrule(lr){2-3}\cmidrule(lr){4-4}\cmidrule(lr){5-5}
& \makecell{\sepehrgh{AutoSigma}\\\sepehrgh{Llama}}
& \makecell{AutoSigma\\OpenAI}
& \makecell{AutoSigma\\Lily}
& \makecell{ChatGPT\\off-the-shelf} \\
\midrule
Mandiant\#1  & \sepehrgh{18} & 21 & 25 & 6 \\
Mandiant\#2  & \sepehrgh{14} & 19 & 22 & 6 \\
Mandiant\#3  & \sepehrgh{30} & 32 & 36 & 5 \\
Mandiant\#4  & \sepehrgh{25} & 28 & 30 & 7 \\
TrendMicro    & \sepehrgh{24} & 24 & 28 & 7 \\
\midrule
\makecell[l]{Avg.}
& \sepehrgh{22}
& 25
& 28 
& 6 \\
\bottomrule
\end{tabular}
}
\end{table}

\sepehrgh{
To further validate the generated rules, we manually extracted the IoCs from the CTI reports and checked whether each IoC appears in at least one of the generated Sigma rules.}
\sepehrgh{
Table~\ref{tab:ioc_coverage_apt41} reports the coverage results using different models. In a Cloud-based LLM environment, \sol achieves near-total coverage of IoCs; on average, it covers 95.6\% and 95.2\% of all IoCs reported in the underlying CTI report. For example, it covers 23 of 25 IoCs (92\%) for Mandiant\#1, 20/21 (95\%) for Mandiant\#2, and 29/32 (91\%) for TrendMicro (APT41). In contrast, ChatGPT-off-the-shelf captures only a few IoCs (\eg 3/25 in Mandiant\#1, 5/32 in APT41), for an average of just 17.6\%. These results indicate that \sol fully incorporates the reported IoCs (several reports reach 100\% coverage), whereas the unguided LLM misses the majority.
%
In Local-based LLMs, \sol covers only about half of the IoCs, coverage ranges from 12/25 (48\%) for Mandiant\#1 to 8/10 (80\%) for Mandiant\#4, with an average of 56.0\%.  Appendix~\ref{sec:local_vs_cloud} provides a discussion on this gap and the difference between local and cloud-based LLM. The trend identified for APT41 holds consistently for the other APTs as well. Across APT28 and APT29, \sol significantly improves IoC and MITRE coverage for both cloud-based and local LLMs, while also achieving higher semantic relevancy than the ChatGPT-off-the-shelf baseline. Detailed per-APT results are reported in Appendix~\ref{sec:avg_apt_results}.}

\begin{table}[!h]
\centering
\caption{IoC-level coverage comparison between APT41 reports and generated Sigma rules.}
\label{tab:ioc_coverage_apt41}
\renewcommand{\arraystretch}{1.12}
\setlength{\tabcolsep}{2pt}
\resizebox{\columnwidth}{!}{
\begin{tabular}{l c cc c c}
\toprule
\multirow{2}{*}{Report}
& \multirow{2}{*}{\makecell{Ground\\Truth}}
& \multicolumn{2}{c}{Cloud-Based LLMs}
& \multicolumn{1}{c}{Local-Based LLMs}
& \multicolumn{1}{c}{Baseline} \\
\cmidrule(lr){3-4}\cmidrule(lr){5-5}\cmidrule(lr){6-6}
& & \makecell{\sepehrgh{AutoSigma}\\\sepehrgh{Llama}} & \makecell{AutoSigma\\OpenAI} 
  & \makecell{AutoSigma\\Lily}
  & \makecell{ChatGPT\\off-the-shelf} \\
\midrule
Mandiant\#1 & 25 & \sepehrgh{23 (92\%)} & 23 (92\%) & 12 (48\%) & 3 (12\%) \\
Mandiant\#2 & 21 & \sepehrgh{19 (90\%)} & 20 (95\%) & 9 (43\%)  & 4 (19\%) \\
Mandiant\#3 & 14 & \sepehrgh{14 (100\%)}& 14 (100\%)& 7 (50\%)  & 3 (21\%) \\
Mandiant\#4 & 10 & \sepehrgh{10 (100\%)}& 10 (100\%)& 8 (80\%)  & 2 (20\%) \\
TrendMicro   & 32 & \sepehrgh{30 (94\%)} & 29 (91\%) & 19 (59\%) & 5 (16\%) \\
\midrule
\makecell[l]{Avg.} 
& -- & \sepehrgh{95.2\%} & 95.6\% & 56.0\% & 17.6\% \\
\bottomrule
\end{tabular}
}
\end{table}

\sepehrgh{We further assessed the technique‐level coverage by extracting the set of \mitre techniques mentioned in each CTI report and comparing it to those appearing in the generated Sigma rules.}
\sepehrgh{To automate this mapping, we employ \textcolor{black}{Threat2Mitre}~\cite{Threat2Mitre2024}, which uses AI-driven language models to process natural-language threat descriptions and link them to the \mitre framework.}

\sepehrgh{We can see in Table~\ref{tab:mitre_coverage_1col_total} that \sol outperforms the baseline in terms of \mitre coverage. With the cloud-based model, \sol OpenAI and \sol Llama achieve an average coverage of 91.2\% and 90.8\% of the techniques extracted from the reports, respectively, whereas ChatGPT-off-the-shelf covers only about 20.2\%, due to the contextual analysis stage of the solution. Using the local LLM yields a lower but still high coverage of 70.8\% for \sol.
These percentages are derived from the per-report counts, for example, in the Mandiant\#3, Threat2Mitre finds 21 techniques in the report, and \sol's cloud-based rules cover 19 of them when using Llama LLM, and 18 when using OpenAI LLM (90\% and 86\% respectively), whereas ChatGPT-off-the-shelf's rules cover only 3 (21\%).
Overall, \sol's Sigma rules retain the vast majority of the techniques mentioned in the reports, demonstrating that the system preserves the original threat intelligence. 
Importantly, \sol also infers new additional techniques beyond those explicitly mentioned in the reports. In total, across the five APT reports, \sol's cloud-based rules introduce 91 new MITRE techniques (an average of 18 additional techniques per report), whereas ChatGPT-off-the-shelf adds only 12 new techniques overall. 
%
This enrichment is driven by \sol's use of external knowledge bases and attack decomposition. By consulting structured \mitre data and decomposing high-level attack behaviors into finer steps, \sol can infer implicit TTPs that the original report did not enumerate. 
This observation is also witnessed across all APTs, reported in Appendix~\ref{sec:avg_apt_results} and Table~\ref{tab:apt_avg_results_baseline}}.

\begin{table}[!h]
\centering
\caption{\mitre-level coverage comparison between APT41 reports and generated Sigma rules.}
\label{tab:mitre_coverage_1col_total}
\renewcommand{\arraystretch}{1.10}
\setlength{\tabcolsep}{3pt}
\resizebox{\columnwidth}{!}{
\begin{tabular}{l c ccc ccc ccc ccc}
\toprule
\multirow{3}{*}{Report}
& \multirow{3}{*}{\makecell{Ground\\Truth}}
& \multicolumn{6}{c}{Cloud-Based LLMs}
& \multicolumn{3}{c}{Local-Based LLMs}
& \multicolumn{3}{c}{Baseline} \\
\cmidrule(lr){3-8}\cmidrule(lr){9-11}\cmidrule(lr){12-14}
& 
& \multicolumn{3}{c}{\makecell{\sepehrgh{AutoSigma}\\\sepehrgh{Llama}}} 
& \multicolumn{3}{c}{\makecell{AutoSigma\\OpenAI}} 
& \multicolumn{3}{c}{\makecell{AutoSigma\\Lily}} 
& \multicolumn{3}{c}{\makecell{ChatGPT\\off-the-shelf}} \\
\cmidrule(lr){3-5}\cmidrule(lr){6-8}\cmidrule(lr){9-11}\cmidrule(lr){12-14}
& 
& Total & Covered & New 
& Total & Covered & New 
& Total & Covered & New 
& Total & Covered & New \\
\midrule
Mandiant\#1 & 8  
& \sepehrgh{16} & \sepehrgh{8 (100\%)} & \sepehrgh{8} 
& 15 & 8 (100\%) & 7 
& 10 & 6 (75\%)  & 4 
& 4 & 2 (25\%)  & 2 \\
Mandiant\#2 & 12 
& \sepehrgh{14} & \sepehrgh{10 (83\%)} & \sepehrgh{4} 
& 17 & 11 (92\%) & 6 
& 9  & 7 (58\%)  & 2 
& 5 & 2 (17\%)  & 3 \\
Mandiant\#3 & 21 
& \sepehrgh{29} & \sepehrgh{19 (90\%)} & \sepehrgh{10} 
& 30 & 18 (86\%) & 12 
& 20 & 14 (67\%) & 6 
& 5  & 3 (14\%)  & 2 \\
Mandiant\#4 & 13 
& \sepehrgh{24} & \sepehrgh{13 (100\%)} & \sepehrgh{11} 
& 20 & 12 (92\%) & 8  
& 16 & 10 (77\%) & 6 
& 6  & 4 (31\%)  & 2 \\
TrendMicro   & 22 
& \sepehrgh{24} & \sepehrgh{18 (81\%)} & \sepehrgh{6}  
& 23 & 19 (86\%) & 4  
& 20 & 17 (77\%) & 3 
& 5  & 3 (14\%)  & 2 \\
\midrule
\makecell[l]{Avg.} 
& -- 
&  & \sepehrgh{90.8\%} & 
&  & 91.2\% & 
&  & 70.8\% & 
&  & 20.2\% & 
\\
\bottomrule
\end{tabular}
}
\end{table}

\sepehrgh{Moreover, we examined the relevancy of generated rules, by checking whether the generated Sigma rules reflect the semantic content of the input threat reports. To quantify this, we compute BERTScore~\cite{zhang2019bertscore} between each CTI report and its corresponding Sigma rules. BERTScore maps tokens to contextual embeddings and computes pairwise cosine similarities, yielding a score $[0 - 1]$ that correlates with semantic alignment. A higher BERTScore indicates that the rule's language captures more of the report's meaning.}

\sepehrgh{Table~\ref{tab:bert_cloud_local_baseline} presents the per-report BERTScores for all models for APT41. Both cloud-based versions of \sol achieve very similar results, with approximately the same average across reports. This shows that \sol is a model-agnostic solution and the variation of the underlying LLM has minimal effect on the relevancy of the generated rules.}
\sepehrgh{
In each APT report, \sol's output consistently achieves the highest relevancy, outperforming the baseline. For example, in Mandiant\#2, the scores are 0.778 (\sol OpenAI ) vs 0.718 (ChatGPT-off-the-shelf). Averaging across all five reports, \sol attains an average BERTScore of 0.764 and 0.765, compared \textcolor{black}{to} 0.718 for ChatGPT-off-the-shelf.}

\begin{table}[!h]
\centering
\caption{BERTScore comparison between the input CTI report and the generated Sigma rules.}
\label{tab:bert_cloud_local_baseline}
\renewcommand{\arraystretch}{1.12}
\setlength{\tabcolsep}{3pt}
\resizebox{\columnwidth}{!}{
\begin{tabular}{l cc c c}
\toprule
\multirow{2}{*}{Report}
& \multicolumn{2}{c}{Cloud-Based LLMs}
& \multicolumn{1}{c}{Local-Based LLMs}
& \multicolumn{1}{c}{Baseline} \\
\cmidrule(lr){2-3}\cmidrule(lr){4-4}\cmidrule(lr){5-5}
& \makecell{\sepehrgh{AutoSigma}\\\sepehrgh{Llama}} & \makecell{AutoSigma\\OpenAI}
& \makecell{AutoSigma\\Lily}
& \makecell{ChatGPT\\off-the-shelf} \\
\midrule
Mandiant\#1 & \sepehrgh{0.766} & 0.762 & 0.722 & 0.709 \\
Mandiant\#2 & \sepehrgh{0.779} & 0.778 & 0.748 & 0.718 \\
Mandiant\#3 & \sepehrgh{0.749} & 0.752 & 0.723 & 0.701 \\
Mandiant\#4 & \sepehrgh{0.771} & 0.775 & 0.745 & 0.729 \\
TrendMicro   & \sepehrgh{0.759} & 0.755 & 0.753 & 0.732 \\
\midrule
\makecell[l]{Avg.} & \sepehrgh{\textbf{0.765}} & \textbf{0.764} & \textbf{0.738} & \textbf{0.718} \\
\bottomrule
\end{tabular}
}
\end{table}

\sol's generated rules include key phrases and contextual details from the report that ChatGPT-off-the-shelf sometimes omits or paraphrases less accurately. We also observed the same pattern when using a local LLM instead of a cloud-based model. Even though absolute scores are slightly lower, \sol using local LLM leads with a mean of 0.738 and 0.718 for ChatGPT-off-the-shelf.
\sepehrgh{\textbf{RQ2.2. Are the rules produced by \sol syntactically valid and executable?}}
We define a generated Sigma rule to be {\em valid} if it is syntactically and operationally correct. In practice, this means the rule must be compiled without errors, and its detection logic should detect at least one relevant malicious event. 
%
\sepehrgh{Table~\ref{tab:validity_apt41_report} shows the generated Sigma rules validity results. We can see that \sol cloud-based, achieves 100\% validity, where every rule generated was valid (100\% across all tested reports).} 
\sepehrgh{In the local-based setting, rule validity is at 76.2\% and it is lower compared to the baseline, because a lighter model is more prone to syntax slips. But, if we consider the number of valid rules that local-based \sol has generated, we have a larger number of rules compared to the baseline. For example, in Mandiant\#4, even though the validity percentage of \sol is lower, it has 25 valid and ready-to-use rules, compared to the 6 valid rules generated by ChatGPT-off-the-shelf. }

\begin{table}[!h]
\centering
\caption{Comparison of rule validity generated by different models across different APT41 reports.}
\label{tab:validity_apt41_report}
\renewcommand{\arraystretch}{1.15}
\setlength{\tabcolsep}{3pt}
\resizebox{\columnwidth}{!}{
\begin{tabular}{l cc c c}
\toprule
\multirow{2}{*}{Report}
& \multicolumn{2}{c}{Cloud-Based LLMs}
& \multicolumn{1}{c}{Local-Based LLMs}
& \multicolumn{1}{c}{Baseline} \\
\cmidrule(lr){2-3}\cmidrule(lr){4-4}\cmidrule(lr){5-5}
& \makecell{\sepehrgh{AutoSigma}\\\sepehrgh{Llama}} & \makecell{AutoSigma\\OpenAI}
& \makecell{AutoSigma\\Lily}
& \makecell{ChatGPT\\off-the-shelf} \\
\midrule
Mandiant\#1 & \sepehrgh{18/18 (100\%)} & 21/21 (100\%) & 18/25 (72\%) & 6/6 (100\%) \\
Mandiant\#2 & \sepehrgh{14/14 (100\%)} & 19/19 (100\%) & 16/22 (73\%) & 6/6 (100\%) \\
Mandiant\#3 & \sepehrgh{30/30 (100\%)} & 32/32 (100\%) & 28/36 (78\%) & 5/5 (100\%) \\
Mandiant\#4 & \sepehrgh{25/25 (100\%)} & 28/28 (100\%) & 25/30 (83\%) & 6/7 (86\%) \\
TrendMicro   & \sepehrgh{24/24 (100\%)} & 24/24 (100\%) & 21/28 (75\%) & 7/7 (100\%) \\
\midrule
\makecell[l]{Avg.}
& \sepehrgh{\textbf{100.0\%}} & \textbf{100.0\%} & \textbf{76.2\%} & \textbf{97.2\%} \\
\bottomrule
\end{tabular}
}
\end{table}

\sepehrgh{\textbf{RQ2.3. Can \sol better adapt to unseen report structures?}}
Adaptability measures whether \sol generates contextually distinct Sigma rules for different input reports, \ie whether semantically different threat reports produce correspondingly different rules rather than generic hallucinations. To assess this, we compute pairwise semantic similarity using BERTScore for (i) the original APT reports, (ii) the Sigma rules generated by \sol, and (iii) the Sigma rules generated by ChatGPT-off-the-shelf. 

\begin{figure}[!h]
    \centering
    \subfigure[APT reports.]{
        \label{fig:RQ5_APT_reports}
        \includegraphics[width=.45\linewidth]{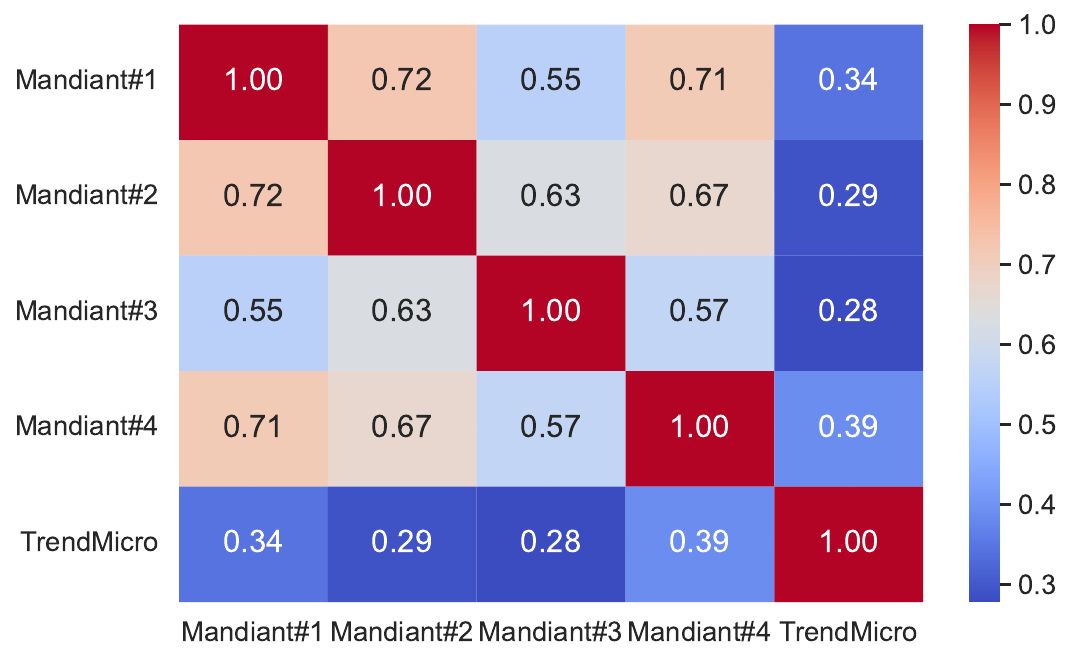}
    }
    \subfigure[\sol OpenAI (Cloud-based).]{
        \label{fig:RQ5_Autosigma}
        \includegraphics[width=.45\linewidth]{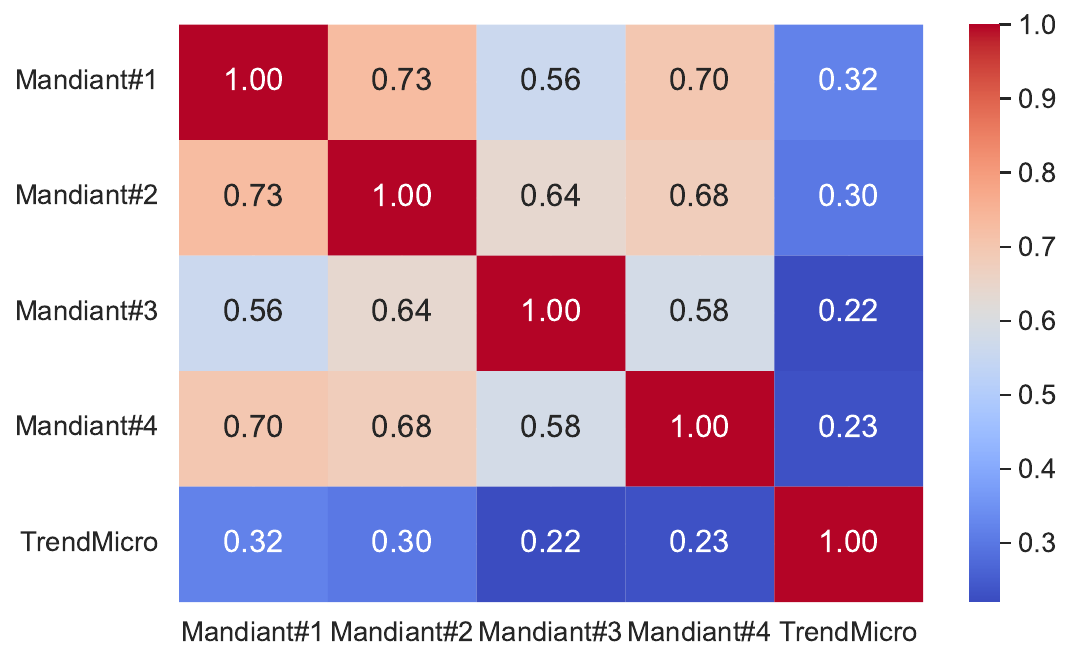}
    }
    \subfigure[\sol Lily (Local-based).]{
        \label{fig:RQ5_Autosigma_local}
        \includegraphics[width=.45\linewidth]{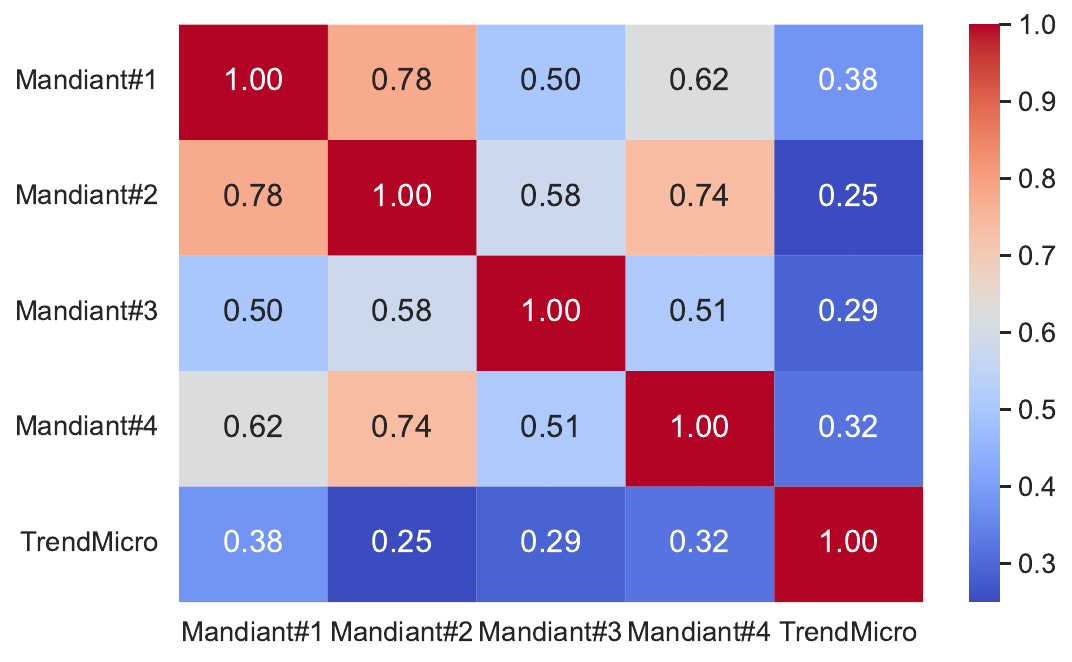}
    }
    \subfigure[ChatGPT-off-the-shelf.]{
        \label{fig:RQ5_ChatGPT_off_the_shelf}
        \includegraphics[width=.45\linewidth]{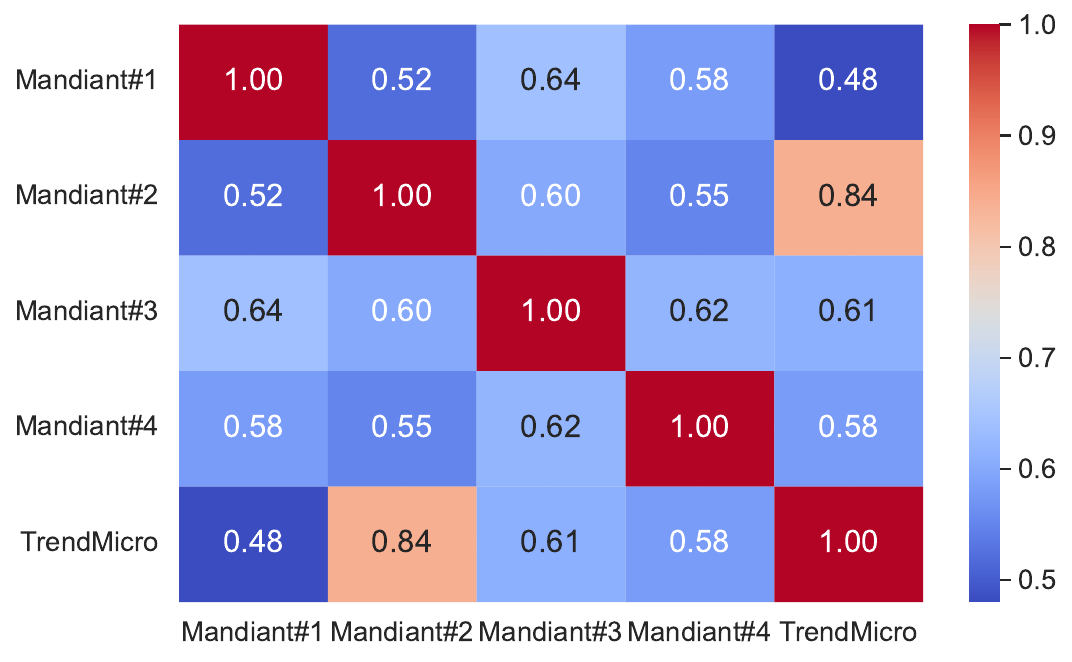}
    }
    \caption{Comparison of semantic similarities in pairs of inputs and pairs of generated Sigma rules by different models across different APT41 reports.}
    \label{fig:RQ5}
\end{figure}

\sepehrgh{Fig. \ref{fig:RQ5_APT_reports} shows the semantic similarity among the original CTI reports. We see a wide range of values ($\approx$0.28-0.72 off the diagonal), confirming that the reports differ substantially in content.
Figs. \ref{fig:RQ5_Autosigma}-\ref{fig:RQ5_ChatGPT_off_the_shelf} show that \sol has much lower off-diagonal similarity overall (means $\approx$0.43-0.50) and values dropping to $\approx$0.17 in some cases. The generated rules from distinct reports are clearly less similar to each other, reflecting the input variability. For example, \sol's cloud-based configurations yield low similarity for pairs of reports that are very different (near 0.17-0.22) and higher similarity only when the original reports are themselves more alike. \sol using a local-based LLM, shows a similar pattern. These results demonstrate that \sol's Sigma rules are contextually tailored, showing that it preserves semantic differences between reports in the output.
By contrast, Fig. \ref{fig:RQ5_ChatGPT_off_the_shelf} shows that ChatGPT-off-the-shelf exhibits uniformly high similarities among rule outputs. Most off-diagonal entries are above $\approx$0.5 (mean $\approx$0.60, with values up to 0.84). This indicates that ChatGPT-off-the-shelf tends to generate very similar (often repetitive) Sigma rules regardless of the input report, failing to differentiate between distinct reports.}

\begin{RQ*}{3}
    How robust is \sol to different design choices (such as the use of LLM-as-a-\textcolor{black}{J}udge, model variation\textcolor{black}{,} and sampling temperature), and how do these choices influence the validity, relevancy\textcolor{black}{,} and stability of generated \textcolor{black}{Sigma} rules\textcolor{black}{?}
\end{RQ*}

\sepehrgh{\textbf{RQ3.1. Ablation study: What is the impact of using an LLM as an automated evaluator on rule validity and rule relevancy?}}
\sepehrgh{In this study, we compare with/without the iterative judge loop on validity. This isolates the benefit of automatic critique-refine cycles. As shown in Table~\ref{tab:validity_apt41_design}, the cloud-based models equipped with the LLM-as-a-\textcolor{black}{J}udge mechanism achieve the highest validity, with 100\% of their generated outputs being valid and directly convertible into SIEM rules. In contrast, models that do not have the LLM-as-a-\textcolor{black}{J}udge component exhibit lower validity ratios. For example, in the TrendMicro report, the \textcolor{black}{cloud}-based \sol (OpenAI) without the LLM-as-a-\textcolor{black}{J}udge generated 21 valid rules out of 24 total rules, whereas using LLM-as-a-\textcolor{black}{J}udge increases this to 100\%.}
\sepehrgh{In the local setting, a similar trend is observed\textcolor{black}{:} the version without LLM-as-a-\textcolor{black}{J}udge achieves an average validity of 47.5\% across all APTs, while integrating the \textcolor{black}{J}udge increases this to 71.7\%, representing an improvement of 24.2\%.}

\begin{table}[!h]
\centering
\caption{Comparison of rule validity by different models across different APT41 reports.}
\label{tab:validity_apt41_design}
\renewcommand{\arraystretch}{1.15}
\setlength{\tabcolsep}{3pt}
\resizebox{\columnwidth}{!}{
\begin{tabular}{l ccccc c}
\toprule
\multirow{2}{*}{Report}
& \multicolumn{4}{c}{Cloud-Based LLMs}
& \multicolumn{2}{c}{Local-Based LLMs} \\
\cmidrule(lr){2-4}\cmidrule(lr){5-7}
& \makecell{\sepehrgh{AutoSigma}\\\sepehrgh{Llama}\\\sepehrgh{(w/ Judge)}} 
& \makecell{\sepehrgh{AutoSigma}\\\sepehrgh{Llama}\\\sepehrgh{(w/o Judge)}} 
& \makecell{AutoSigma\\OpenAI\\(w/ Judge)} 
& \makecell{AutoSigma\\OpenAI\\(w/o Judge)} 
& \makecell{AutoSigma\\Lily\\(w/ Judge)} 
& \makecell{AutoSigma\\Lily\\(w/o Judge)} \\
\midrule
Mandiant\#1 & \sepehrgh{18/18 (100\%)} & \sepehrgh{16/18 (89\%)} & 21/21 (100\%) & 18/21 (86\%) & 18/25 (72\%) & 13/25 (52\%) \\
Mandiant\#2 & \sepehrgh{14/14 (100\%)} & \sepehrgh{12/14 (86\%)} & 19/19 (100\%) & 17/19 (89\%) & 16/22 (73\%) & 10/22 (45\%) \\
Mandiant\#3 & \sepehrgh{30/30 (100\%)} & \sepehrgh{24/30 (80\%)} & 32/32 (100\%) & 26/32 (81\%) & 28/36 (78\%) & 20/36 (56\%) \\
Mandiant\#4 & \sepehrgh{25/25 (100\%)} & \sepehrgh{21/25 (84\%)} & 28/28 (100\%) & 26/28 (93\%) & 25/30 (83\%) & 16/30 (53\%) \\
TrendMicro   & \sepehrgh{24/24 (100\%)} & \sepehrgh{21/24 (88\%)} & 24/24 (100\%) & 21/24 (88\%) & 21/28 (75\%) & 12/28 (43\%) \\
\midrule
\makecell[l]{Avg.} 
& \sepehrgh{\textbf{100.0\%}} & \sepehrgh{\textbf{85.4\%}} & \textbf{100.0\%} & \textbf{87.4\%} & \textbf{76.2\%} & \textbf{49.8\%} \\
\bottomrule
\end{tabular}
}
\end{table}

\sepehrgh{\textbf{RQ3.2. Design choice study: How does the sampling temperature influence the balance between rule diversity and hallucination frequency?}}
%
\sepehrgh{Hallucination is a major challenge in LLM-based solutions. Although \sol uses LLM-as-a-\textcolor{black}{J}udge to mitigate the risk of \textcolor{black}{hallucination}, there is currently no standardized metric for quantifying hallucinated content in this context. Therefore, we \textcolor{black}{analyze} behaviors that affect the likelihood of hallucination and present the measures taken in \sol to reduce this likelihood.}

\begin{figure}[t]
    \centering
    \includegraphics[width=.65\linewidth]{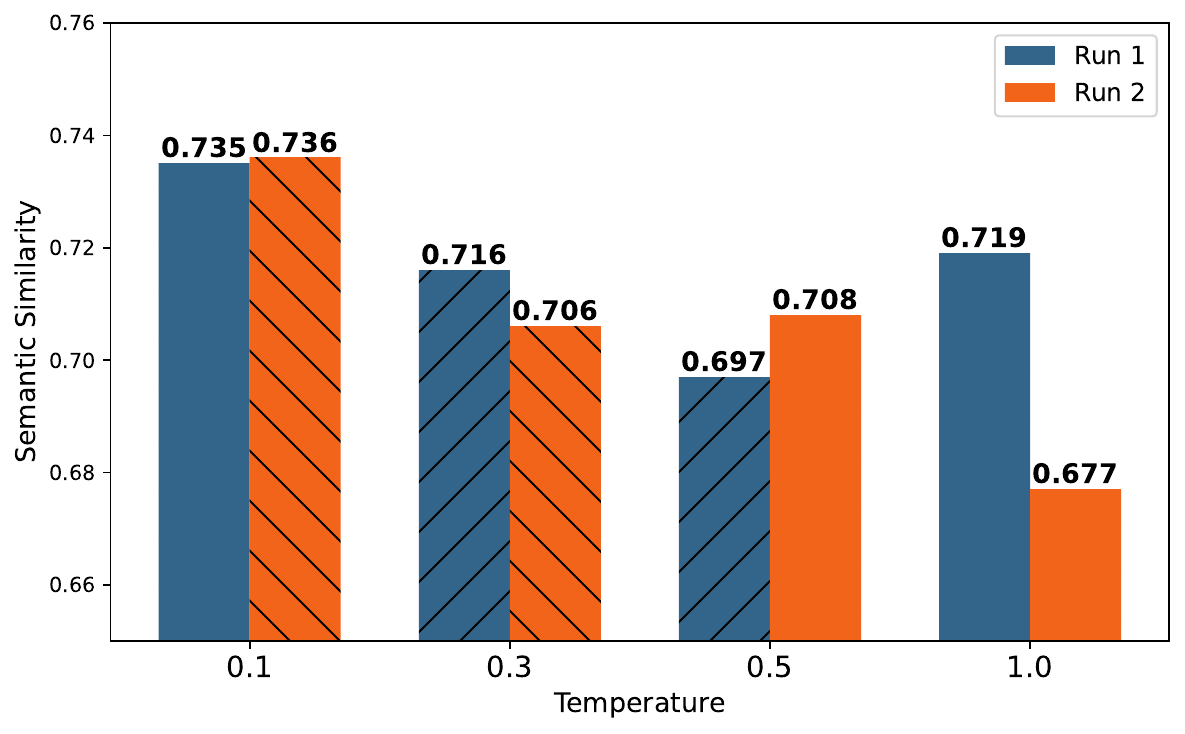}
    \caption{Effect of temperature on semantic similarity.}
    \label{fig:temp_fig}
\end{figure}

\sepehrgh{Temperature~\cite{holtzman2020curiouscaseneuraltext} is a key controllable parameter in LLM sampling that controls randomness and diversity. Higher temperatures tend to increase linguistic variation and “creativity” but also introduce instability and hallucinated logic. To evaluate its impact, we conducted an experiment using one representative APT report and ran \sol twice per temperature value $T = \{0.1, 0.3, 0.5, 1.0\}$. For each run, we computed the semantic similarity using BERTScore between the generated rules and the original CTI report. As shown in Fig.~\ref{fig:temp_fig}, the outputs at $T = 0.1$ remain consistent across runs, while higher temperatures cause semantic divergence. This confirms that maintaining a low temperature (0.1 in our main experiments) provides stable, low-variance results and effectively suppresses hallucinated content.}

\begin{figure}[t]
    \centering
    \includegraphics[width=.65\linewidth]{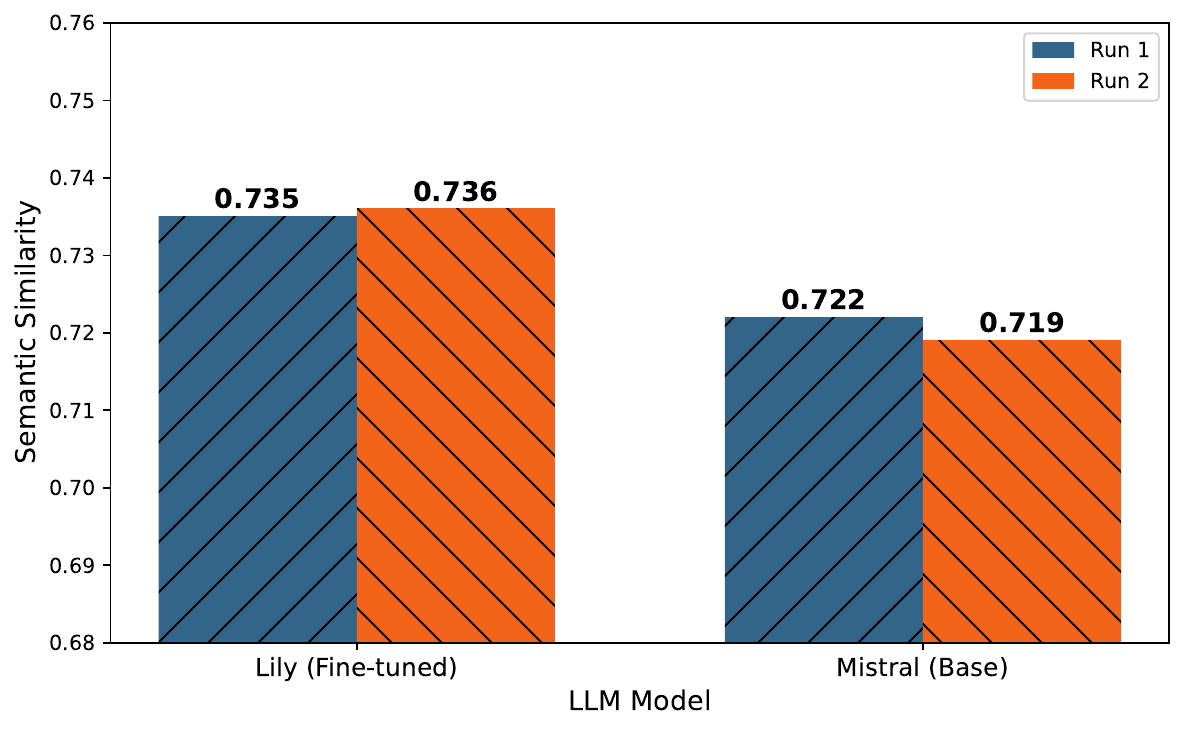}
    \caption{Effect of model fine-tuning on hallucination.}
    \label{fig:fine_tune_fig}
\end{figure}

\sepehrgh{\textbf{RQ3.3. Does domain-specific fine-tuning reduce hallucination and improve coverage compared to prompt-only generation?}}
\sepehrgh{Fine-tuning the base LLM with domain-specific data can further improve the factual grounding and consistency of generated rules. To validate this, we compared two variants of \sol using the same temperature setting ($T=0.1$): one with the cybersecurity fine-tuned model \textit{Lily-Cybersecurity-7B-v0.2} and another using a generic Mistral LLM. As shown in Fig.~\ref{fig:fine_tune_fig}, the fine-tuned Lily model achieves consistently higher semantic similarity across runs, confirming its reduced hallucination behavior. This observation aligns with prior work such as CyberPalAI~\cite{levi2024cyberpalaiempoweringllmsexpertdriven}.}

\textbf{RQ3.4. How does \sol perform with noisy data?}
\sepehr{Although our framework assumes that CTI reports are reliable and that CTI poisoning or manipulation is out of scope, in practice external CTI should be considered untrusted. To assess the robustness of \sol under degraded input conditions, we conduct an experiment where controlled Gaussian noise~\cite{bishop2006pattern} is injected into the CTI reports at different contamination rates. The noise is applied at the text level using character-level perturbations, and the resulting outputs are compared against those generated from the original clean reports using similarity metrics.}
\begin{figure}[!h]
    \centering
    \includegraphics[width=.9\linewidth]{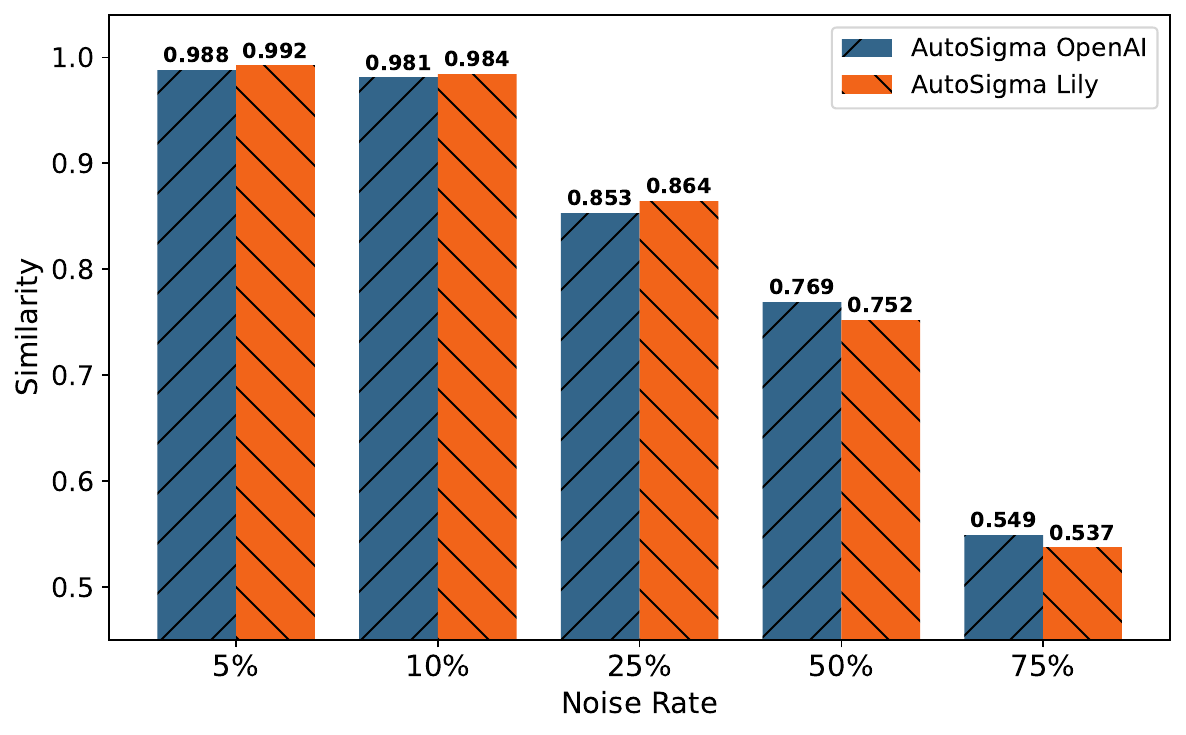}
    \caption{Effect of noisy data.}
    \label{fig:noise_robustness}
\end{figure}
\sepehr{The results are shown in Fig.~\ref{fig:noise_robustness}. As expected, the similarity decreases as the contamination rate increases. Up to a 10\% contamination rate, \sol maintains a high similarity above 0.98, indicating strong robustness to minor perturbations. At 25\% noise, the similarity drops to approximately 0.85, reflecting moderate degradation. Beyond this point, the performance declines more significantly, reaching 0.75 at 50\% and 0.53 at 75\% contamination. These results demonstrate that while \sol is resilient to low levels of noise, substantial corruption in the input can negatively impact the quality of the generated rules.
It is important to note that \sol is designed to operate on validated CTI reports typically used by SOC teams. Handling adversarial manipulation or heavily corrupted CTI inputs is beyond the scope of this work and is left for future investigation.}

\section{Discussion}
The evaluation of \sol highlights its strong performance in generating high-quality, deployable Sigma rules. However, beyond reporting numerical metrics, it is essential to critically assess the solution’s design and identify areas where improvements can be made.
A promising direction for enhancing robustness involves combining generative reasoning with rule-based validation logic, such as leveraging structured grammars, detection logic templates, and constraints informed by known adversary behaviors. This would enable more principled rule synthesis that does not depend solely on LLM fluency. Additionally, the current rule generation loop is generative at every iteration. Moving toward a guided refinement process, where intermediate rule candidates are evaluated not only by another model but also by static analyzers or predefined criteria, could reduce hallucinations and enforce stronger alignment with security semantics.
A separate concern is the execution time of the system. Because \sol operates as a multi-stage pipeline with enrichment, template retrieval, and iterative validation, the overall latency can be non-trivial. While this is acceptable for offline threat hunting use cases, where rules are generated as part of periodic analysis, it becomes a limitation in time-sensitive environments. \textcolor{black}{Although the pipeline was not designed for real-time deployment}---\sepehrgh{it is designed for \textcolor{black}{offline} deployment}---\textcolor{black}{improving the} efficiency of the validation loop and introducing caching or early-exit strategies could significantly reduce processing delays and make the system more responsive. Despite these performance trade-offs, it is worth emphasizing that the rule synthesis process only needs to run once per report and does not interfere with real-time alert handling, making this overhead manageable in most hunting workflows.
\sepehr{A further limitation relates to the training data of the underlying large language models. For cloud-based models such as OpenAI GPT-4o and LLaMA, the exact composition of the training data is not publicly disclosed, and therefore we cannot guarantee that the CTI reports used in our evaluation were not part of their pretraining data. While these models are general-purpose and not specifically trained for CTI-to-Sigma rule generation tasks, the possibility of prior exposure cannot be fully ruled out. Similarly, for the local model (Lily), although it is fine-tuned on cybersecurity data, the exact sources of this data are not fully specified. This limitation is inherent to current LLM-based systems and should be considered when interpreting the results.}

\section{Conclusion}
In this paper, we introduced \sol, a fully automated framework for generating actionable Sigma rules from unstructured CTI reports. Through the integration of knowledge-based enrichment, template-guided rule generation, and a structured multi-phase pipeline, \sol produces contextually relevant and semantically accurate Sigma rules that can be used for hunting activities. 
By leveraging external threat knowledge, aligning with existing detection templates, and enforcing rule correctness through iterative validation, \sol offers a robust and effective mechanism for streamlining threat detection and enhancing security workflows.
Our evaluation using public CTI reports demonstrated that \sol achieves high validity in rule generation, consistently outperforming baseline models, and requiring minimal manual intervention.
%
Future work will involve validating \sol across broader, more heterogeneous datasets and in operational environments involving live alert streams.

\section*{Acknowledgments}
The authors used Generative AI (\ie GPT-4o) solely for grammar and language refinement.

\bibliographystyle{IEEEtran}
\bibliography{Ref}

\appendices

\section{Example of Outputs}
\label{sec:output_example}

\begin{figure}[!ht]
    \centering
    \subfigure[An example of attack scenario enrichment output.]{
        \label{fig:Attack_enrichment_json}
        \includegraphics[width=0.44\linewidth]{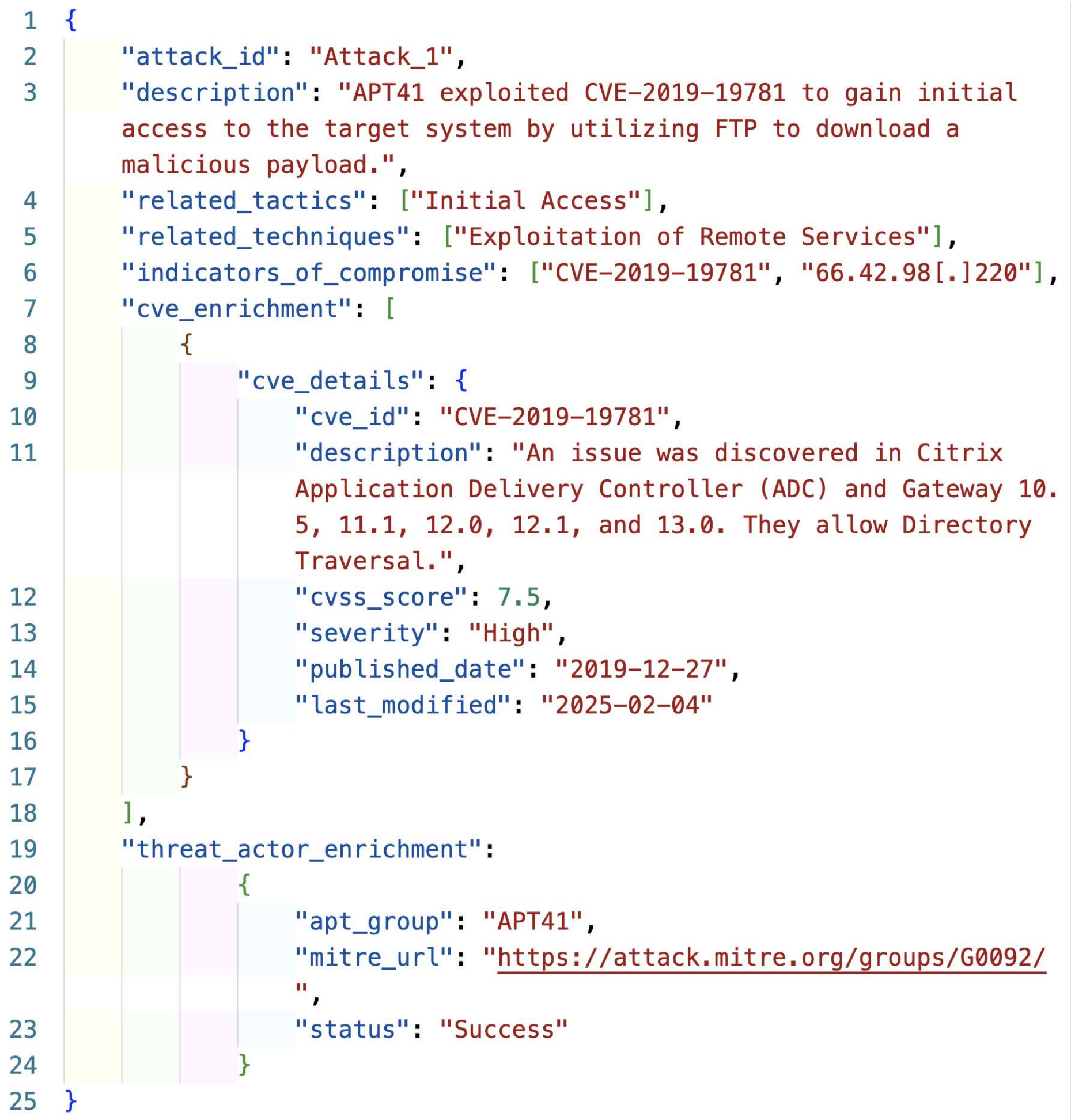}
    }
    \subfigure[An example of attack step decomposition output.]{
        \label{fig:attack_simplification_json}
        \includegraphics[width=0.44\linewidth]{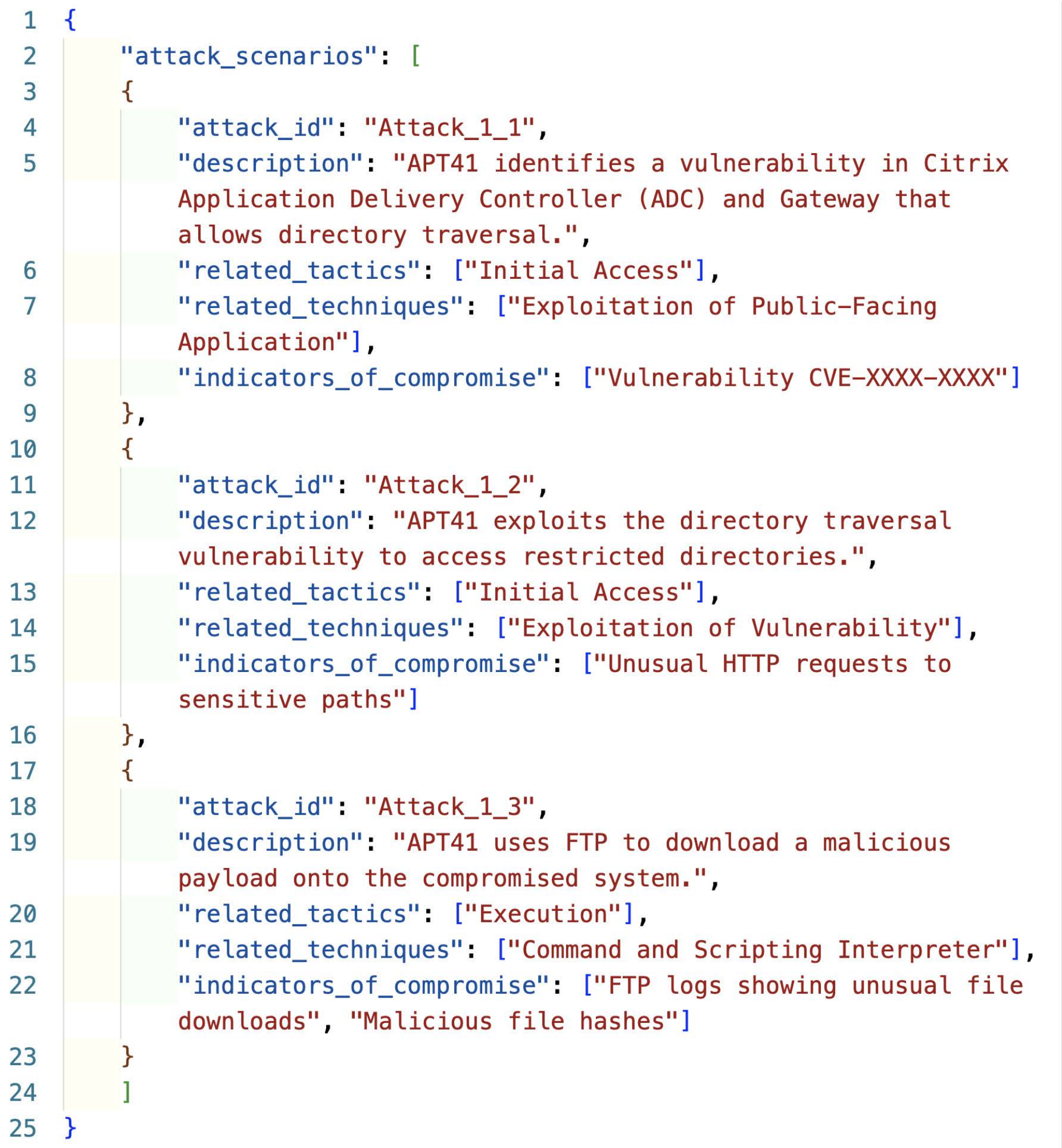}
    }
    \caption{Outputs of attack scenario enrichment and step decomposition.}
    \label{fig:json_example}
\end{figure}

Fig.~\ref{fig:json_example} shows an example of attack scenario enrichment and attack step decomposition. In Fig.~\ref{fig:Attack_enrichment_json}, we observe the output of the enrichment phase, which includes extracted entities such as the CVE (lines 9-17), threat actor (lines 19-24), and contextual description. In Fig.~\ref{fig:attack_simplification_json}, the enriched scenario is decomposed into three smaller attack steps: the first involves the identification of a vulnerability in Citrix ADC (Initial Access), the second describes the actual exploitation attempt using that vulnerability, and the third captures the follow-up action of downloading a malicious payload using FTP (Execution phase). Each of these steps is now ready to be handled separately by the rule generation module.

\section{LLM Prompting}
\label{sec:prompts}
\begin{figure}[!h]
    \centering
    \subfigure[NER Extractor Component.]{
        \label{fig:prompt_example_ner}
        \includegraphics[width=0.44\linewidth]{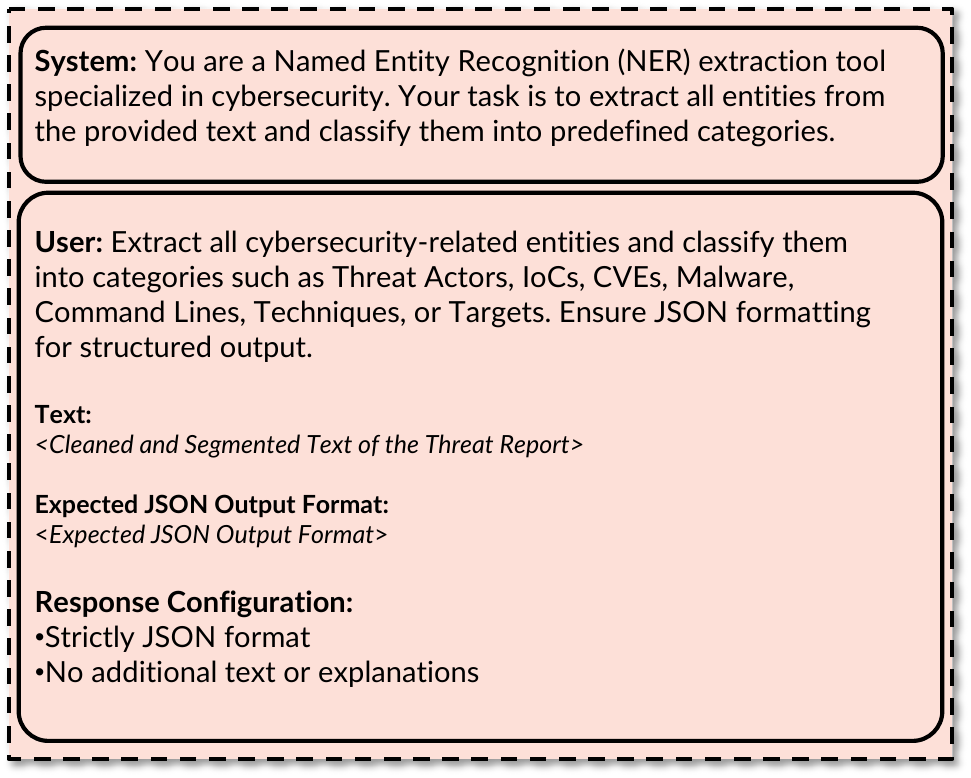}
    }
    \subfigure[Attack Scenario Extractor Component.]{ 
        \label{fig:prompt_example_attack}
        \includegraphics[width=0.44\linewidth]{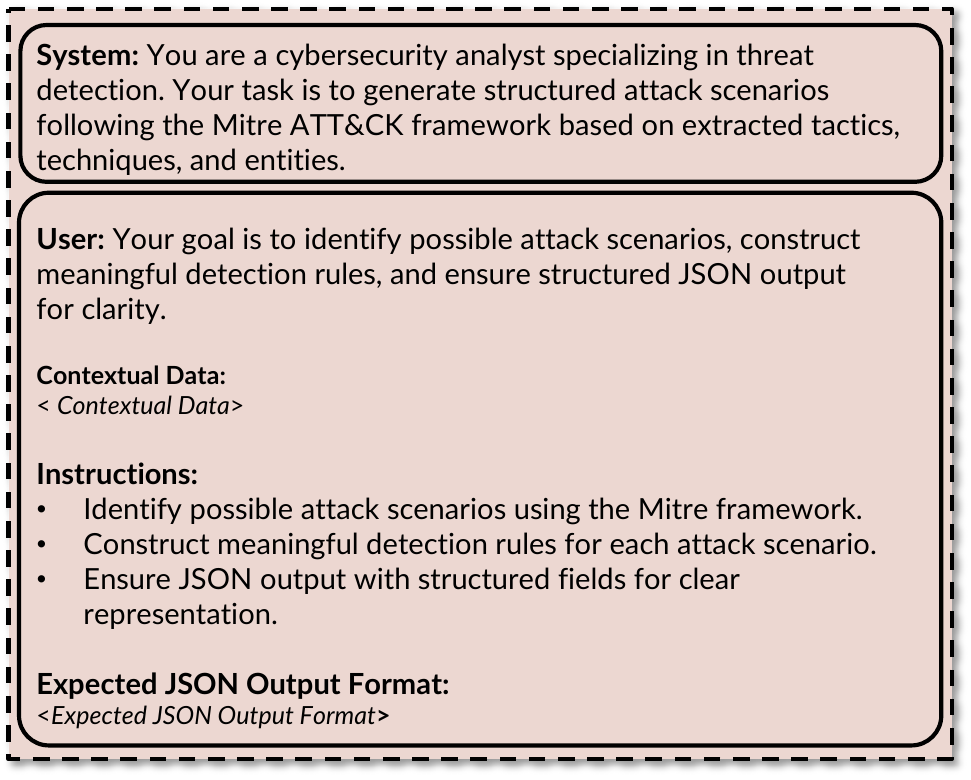}
    }
    \caption{Example of prompts used by \sol.}
    \label{fig:prompt_example}
\end{figure}

Basic prompts such as "extract NERs from a given text" proved inadequate for cybersecurity tasks, which demand higher specificity and structure. To address this, we employed prompt engineering techniques that included explicit task instructions, relevant examples, and clearly defined output formats. An illustration of our \sepehrgh{structured prompt engineering} is shown in Fig.~\ref{fig:prompt_example}.

\begin{figure}[!t]
    \centering
    \subfigure[Prompt used for Sigma rule generator.]{
        \label{fig:prompt_sigma_enricher}
        \includegraphics[width=0.44\linewidth]{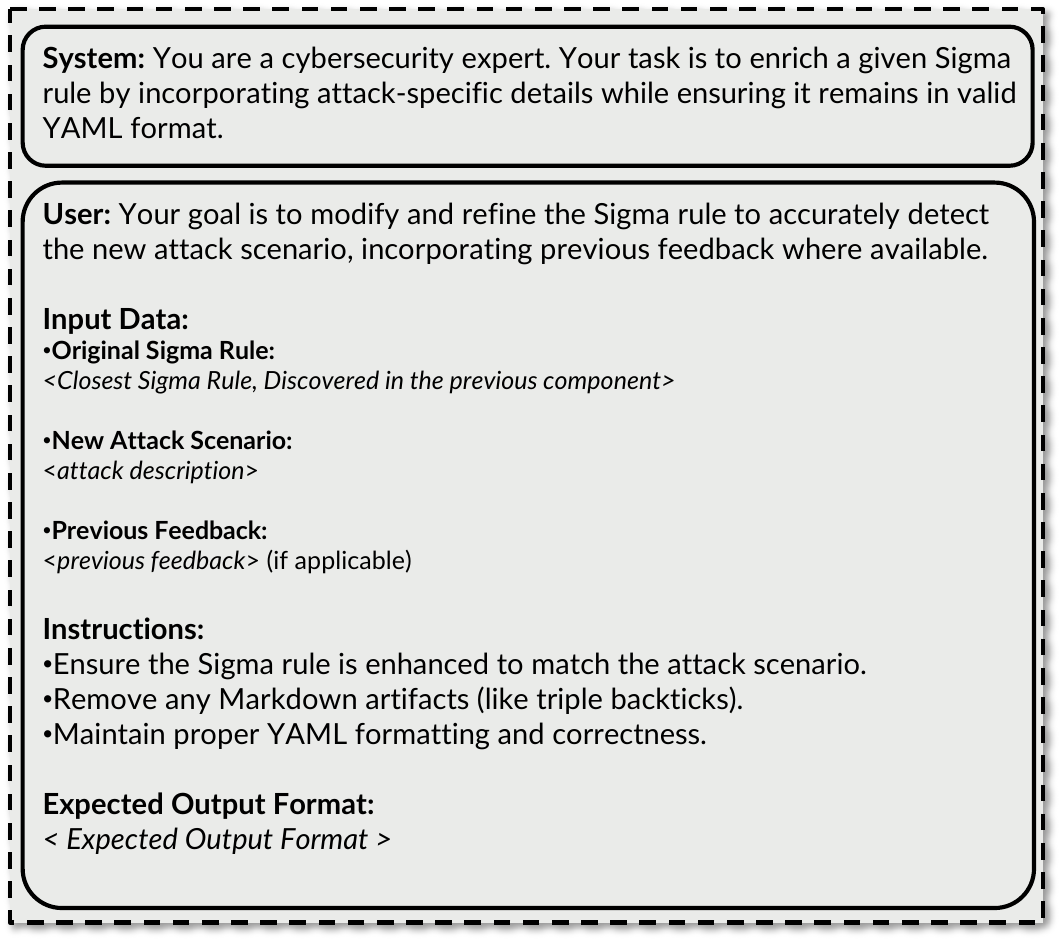}
    }
    \subfigure[Prompt used for Sigma rule validator.]{ 
        \label{fig:prompt_sigma_validator}
        \includegraphics[width=0.44\linewidth]{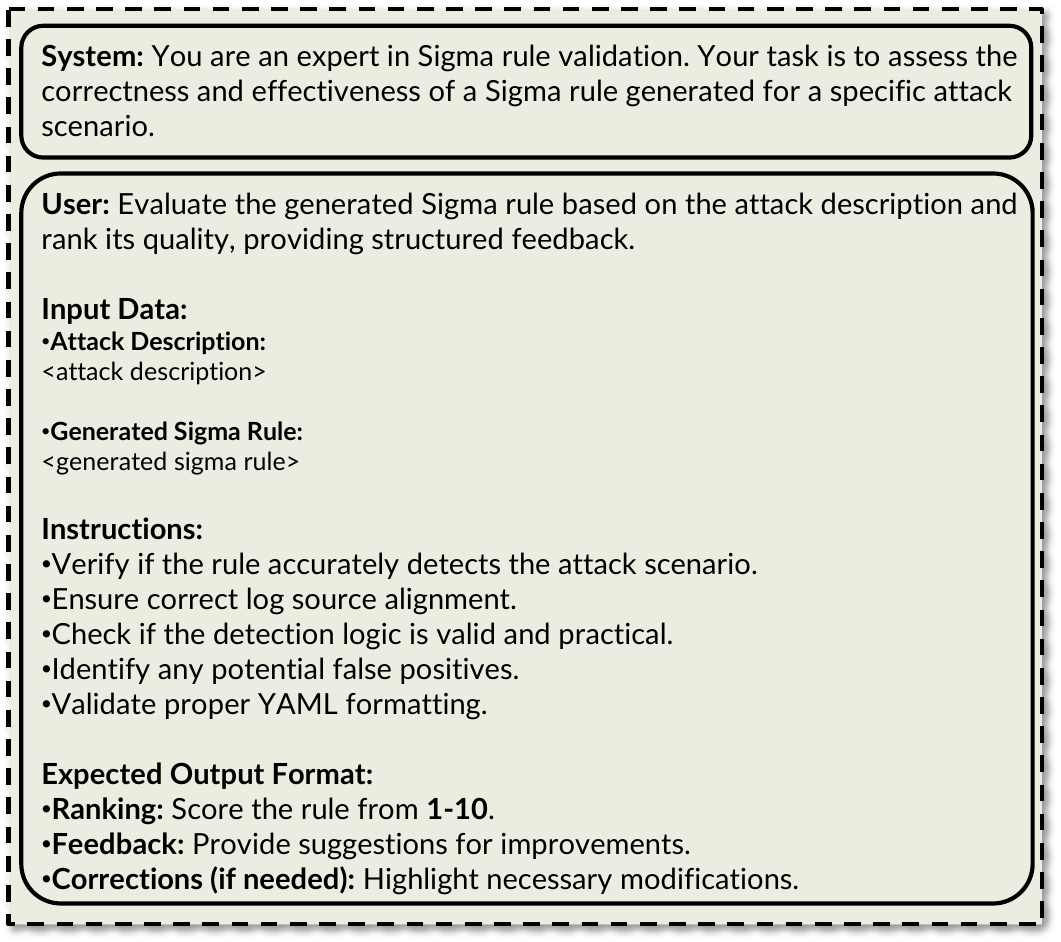}
    }
    \caption{LLM-as-a-Judge prompts.}
    \label{fig:prompt_LLM_Judge}
\end{figure}

\subhead{LLM-as-a-Judge Loop}
To ensure the quality of the generated Sigma rules, \sol employs an LLM-as-a-Judge loop in which a generator and a validator model work iteratively. The generator produces an initial Sigma rule (as shown in Fig.~\ref{fig:prompt_sigma_enricher}), a separate LLM, used as the validator, then evaluates the generated rule against predefined criteria, including correctness, completeness, and Sigma syntax compliance (Fig.~\ref{fig:prompt_sigma_validator}). If the validator's score falls below a fixed threshold (8/10 across all dimensions), its feedback is returned to the generator, and the refinement loop continues until the quality criteria are met. \sepehrgh{Furthermore, as illustrated in Fig.~\ref{fig:prompt_LLM_Judge}, the prompts used for LLM-as-a-\textcolor{black}{J}udge \textcolor{black}{are} adaptive and dynamically constructed.}

\section{Proof of Concept (PoC)}
\label{sec:PoC}

\begin{figure*}[!ht]
    \centering
    \includegraphics[width=.95\linewidth]{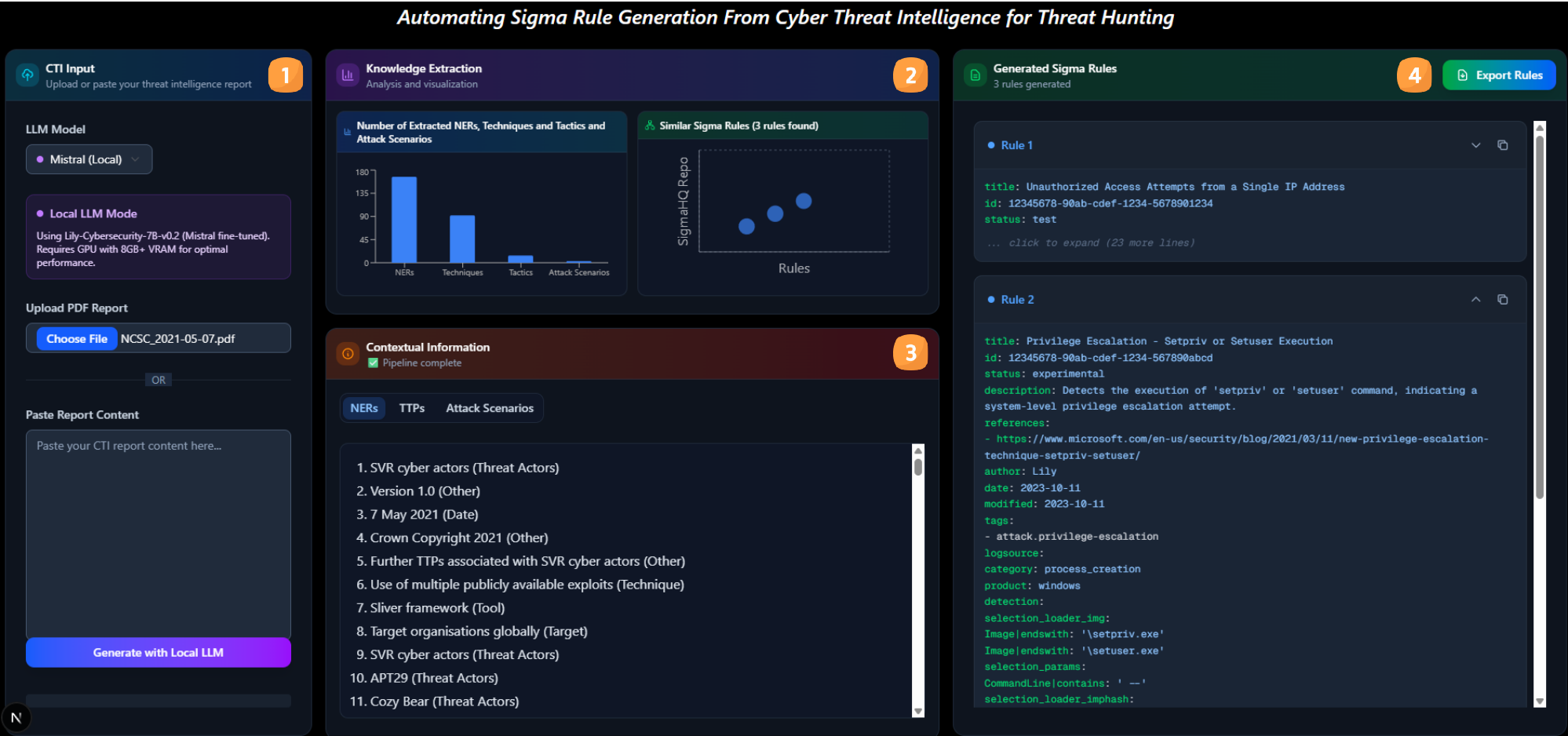}
    \caption{Screenshot of the \sol PoC interface.}
    \label{fig:AutoSigma_PoC}
\end{figure*}

\sepehrgh{
Fig.~\ref{fig:AutoSigma_PoC} presents the Proof of Concept interface we developed for \sol, highlighting the end-to-end workflow of the \sol. The UI is structured into four main sections, visually indicated by blue rounded-square icons:}

\begin{itemize}[label={}]
    \item \squircle{1}\sepehrgh{\textbf{Input Area:} Users select the LLM model (local or cloud-based) and upload or paste the threat report to initiate analysis.}
    \item \squircle{2}\sepehrgh{\textbf{Knowledge Extraction \& Similar Rule Discovery:} The left panel visualizes key statistics such as the total number of extracted Named Entities (NERs), MITRE ATT\&CK techniques, tactics, and attack tests. The right panel displays a bubble chart summarizing similar Sigma rules identified from the SigmaHQ repository.}
    \item \squircle{3}\sepehrgh{\textbf{Contextual Information:} This area shows the detailed breakdown of contextual information extracted by \sol, including recognized NERs, ATT\&CK techniques, tactics, and the reconstructed attack scenario.}
    \item \squircle{4}\sepehrgh{\textbf{Outputs and Rules:} The final output section displays the generated Sigma detection rules and allows users to review or export them.}
\end{itemize}

\begin{table*}[!h]
    \centering
    \caption{Summary of CTI reports used in the evaluation.}
    \label{tab:reports}
    \resizebox{\linewidth}{!}{
    \begin{tabularx}{\linewidth}{l|X|X|c|c|c} 
        \hline
        \textbf{Report} & \textbf{Target Industries} & \textbf{Main TTPs} & \textbf{Detected} & \textbf{Published} & \textbf{Pages} \\ 
        \hline
        \multicolumn{6}{c}{\textbf{APT41}} \\
        \hline
        Mandiant\#1~\cite{mandiant2024} & Banking , Finance, Government, High-tech & PowerShell usage, Cobalt deployment, FTP for payload delivery & Jan-Mar 2020 & 2020-03-25 & 13 \\ 
        Mandiant\#2~\cite{googleblog2023} & Gaming, Healthcare, High-tech & Exploiting vulnerabilities, Web shells, Backdoors & Apr 2019 & 2019-08-19 & 9 \\ 
        Mandiant\#3~\cite{googleblog2024} & Telecoms, News/Media, Software firms & Backdoors, Reconnaissance, Credential theft & Since 2012 & 2019-08-07 & 68 \\ 
        Mandiant\#4~\cite{googleblogAPT41states} & U.S. State Gov, Insurance, Telecom & SQL injection, Zero-day exploits, DNS manipulation & May 2021-Feb 2022 & 2022-03-08 & 9 \\ 
        TrendMicro~\cite{trendmicro2021earth} & Enterprises, Government, Airline, Publishing & StealthVector, Backdoors, DNS manipulation & Jul 2020 & 2021-08-24 & 55 \\ 
        \hline
        \multicolumn{6}{c}{\textbf{APT28}} \\
        \hline
        ANSSI~\cite{ANSSI2023} & Gov., universities, think tanks (France) & CVE-2023-23397, phishing, compromised edge devices & H2 2021-2023 & 2023-10-26 & 7 \\
        ClearSky~\cite{ClearSky2024} & Media, public info outlets (US, EU, Israel) & Disinformation, fake news infra reuse & 2023-2024 & 2024-02-22 & 29 \\
        Insikt\#1~\cite{Insikt2023a} & Ukrainian government & Roundcube/Outlook phishing (CVE-2023-23397), credential theft & 2021-2023 & 2023-06-20 & 17 \\
        Logpoint~\cite{Logpoint2024} & NATO-aligned gov./defense/media & GooseEgg exploit, Print Spooler abuse, LOLBINs & 2023-2024 & 2024-04-05 & 29 \\
        Maverits~\cite{Maverits2023} & Gov., foreign affairs (Ukraine, Europe) & Backdoors, LOLBINs, cybercrime collaboration & 2022-2023 & 2023-01-27 & 32 \\
        \hline
        \multicolumn{6}{c}{\textbf{APT29}} \\
        \hline
        CrowdStrike~\cite{crowdstrikestellarparticle2003} & Foreign ministries, NATO governments & Credential phishing, malware loader & 2021-2022 & 2022-01-27 & 21 \\
        HKUK~\cite{HKUK2023} & Multiple UK sectors incl. academia & Password spraying, VPN compromise, supply chain & 2021-2023 & 2023-11-14 & 12 \\
        Insikt~\cite{recordedfuturebluebravo2024} & Think tanks, foreign policy institutions & Credential theft via phishing themes & 2022-2023 & 2023-07-27 & 19 \\
        Mandiant\#5~\cite{mandiantapt292018} & U.S. and European gov. orgs & Credential theft, C2 tunneling, phishing kits & 2016-2018 & 2018-11-19 & 17 \\
        NCSC~\cite{ncsc2021} & Defense, law, gov. entities & Credential spraying, Exchange exploit & 2019-2021 & 2021-05-07 & 7 \\
        \hline
    \end{tabularx}
    }
\end{table*}

\section{CTI Reports Summary}
\label{sec:apt41_summary}
Table~\ref{tab:reports} provides a summary of the used CTI reports per APT in our evaluation.
\sepehrgh{To reflect the diversity and realism of operational CTI, we selected reports from multiple security vendors, written at different levels of technical depth, and varying significantly in both length and structure. These reports mix multiple forms of information, including high-level executive summaries, detailed technical analyses, command line snippets, screenshots, figures, and narrative descriptions of attacker behavior. Using such a varied corpus demonstrates that \sol can operate effectively across heterogeneous and unstructured report formats, regardless of whether the content is tactical, technical, or strategic.
}

\section{Local LLMs Versus Cloud LLMs}
\label{sec:local_vs_cloud}
While our evaluation compared both cloud-based and local LLM deployments, it is important to reiterate that \sol was designed as a pipeline, not as a benchmark for individual LLMs. The performance gaps we observed between the two settings reflect differences in model capabilities at specific stages of the pipeline rather than shortcomings in the design. In fact, our results suggest that each class of model brings unique strengths to different parts of the system.
The local LLM used in our experiments , a fine-tuned variant of Mistral with a cybersecurity-specific training corpus ,  performed particularly well during the early analytical stages. Tasks such as NER and TTP extraction, and attack scenario extraction benefited from the model's domain alignment. In these steps, the local model demonstrated strong precision and contextual understanding, likely due to the security-specific patterns it had learned during fine-tuning.
However, the final stages of the pipeline, especially the generation of Sigma rules, placed greater demands on generative capacity, instruction adherence, and formatting consistency. Here, the cloud-based models exhibited a clear advantage. Backed by significantly larger parameter counts and broader pretraining data, these cloud-based models consistently produced more well-formed, coherent, and complete Sigma rules. Because our validation and scoring processes take place after generation, the strength of the initial drafts plays a decisive role in final outcomes. This explains why, even though the local model handled upstream analysis well, the API models slightly outperformed it in downstream metrics like rule validity and semantic alignment.
This analysis also clarifies that the observed performance delta is not due to the pipeline failing on local models, but to generative limitations that the generator and validator alone cannot fully overcome. It also opens a path for hybrid approaches, such as using a local model for extraction and an API model for generation, offering future work opportunities to balance performance and deployment constraints more effectively.
\sepehrgh{\textcolor{black}{It} is also worth noting that, as discussed earlier, cloud-based LLMs introduce a small operational cost, while removing the need for computational resources. \sepehr{In our experiments, the cloud-based versions of \sol had an estimated cost of \$0.0015 - \$0.003 per run, depending on the number of iterations the LLM-as-a-\textcolor{black}{J}udge loop takes.} Since these models require no dedicated GPU resources or local model hosting, \textcolor{black}{this makes them} convenient for lightweight deployment. Conversely, local LLMs are free to use, yet demand sufficient hardware capacity, typically GPUs, to deploy. These trade-offs highlight an important design consideration for practitioners.}


\sepehrgh{\section{LLM Specifications}}
\label{sec:llm_specification}

\sepehrgh{For our experiments, we used three LLMs across both cloud-based and local configurations, selected to capture variations in architecture, parameter scale, and domain specialization. All models were executed with a fixed temperature of $0.1$ and a maximum generation length of 512 tokens to ensure deterministic and reproducible outputs.}

\begin{itemize}
    \item \sepehrgh{\textbf{GPT-4o-mini}: a general-purpose LLM designed for tasks such as reasoning, summarization, and text generation. It serves as a representative commercial closed-weight model.}
    \item \sepehrgh{\textbf{Llama 3.3-70B}: an open-weight, large-scale model with 70 billion parameters, capable of multi-turn reasoning and contextual comprehension. It is used to assess performance consistency across model families under identical prompting conditions.}
    \item \sepehrgh{\textbf{Lily-Cybersecurity-7B-v0.2}: a fine-tuned version of Mistral-7B adapted to the cybersecurity \textcolor{black}{domain}, with 22,000 hand-crafted cybersecurity and hacking-related data pairs. The dataset was then run through an LLM to provide additional context, personality, and styling to the outputs.}
\end{itemize}

\sepehrgh{\section{Average Evaluation \textcolor{black}{Across} Multiple APTs}}
\label{sec:avg_apt_results}

\sepehrgh{
As shown in Table~\ref{tab:apt_avg_results_baseline}, the trend observed for APT41 holds consistently for APT28 and APT29 as well. \sol significantly improves IoC coverage using both cloud-based and local LLMs. With cloud-based LLMs, the average IoC coverage increases from 20.1\% (ChatGPT-off-the-shelf baseline) to 90.4\%. Similarly, in the local LLM setting, \sol improves the average IoC coverage from 20.1\% to 60.7\%, corresponding to an increase of 40.6 percentage points.
}
\sepehrgh{
This observation is also consistent with our MITRE ATT\&CK coverage results (see Table~\ref{tab:mitre_coverage_1col_total} in the main text), demonstrating that \sol not only matches the report-derived technique coverage, but also significantly broadens it, yielding a more comprehensive Sigma rule set that covers more techniques across different APT campaigns.
}
\sepehrgh{
The same pattern appears when considering semantic relevancy. For complex reports in APT28 and APT29, the average relevancy of \sol is high (\ie 0.816 and 0.814), while ChatGPT-off-the-shelf lags behind (\ie 0.711 and 0.690). This confirms that the rules generated by \sol capture the critical context and tactics of each campaign rather than producing generic or poorly aligned detections.
}

\begin{table}[H]
\centering
\caption{Average results of all APTs across models.}
\label{tab:apt_avg_results_baseline}
\renewcommand{\arraystretch}{1.2}
\resizebox{\columnwidth}{!}{
\begin{tabular}{l|c|c|c|c|c}
\hline
\noalign{\vskip 2pt}
\textbf{APT} & \textbf{Model} 
& \makecell{IoC\\Coverage}
& \makecell{MITRE\\Coverage}
& \makecell{Relevancy}
& \makecell{Validity} \\
\hline

\multirow{4}{*}{APT41} 
& \makecell{\sepehrgh{\sol Llama (Cloud-based)}}   & \sepehrgh{95.2} & \sepehrgh{90.8} & \sepehrgh{76.5} & \sepehrgh{100} \\
& \makecell{\sol OpenAI (Cloud-based)}            & 95.6 & 91.2 & 76.4 & 100 \\
& \makecell{\sol Lily (Local-based)}               & 56.0 & 70.8 & 73.8 & 76 \\
& \makecell{ChatGPT-off-the-shelf}                 & 17.6 & 20.2 & 71.8 & 97 \\
\hline

\multirow{4}{*}{APT29} 
& \makecell{\sepehrgh{\sol Llama (Cloud-based)}}   & \sepehrgh{86.0} & \sepehrgh{91.6} & \sepehrgh{81.6} & \sepehrgh{100} \\
& \makecell{\sol OpenAI (Cloud-based)}            & 87.0 & 91.6 & 81.4 & 100 \\
& \makecell{\sol Lily (Local-based)}               & 62.4 & 63.6 & 76.1 & 75 \\
& \makecell{ChatGPT-off-the-shelf}                 & 28.0 & 29.4 & 69.0 & 93 \\
\hline

\multirow{4}{*}{APT28} 
& \makecell{\sepehrgh{\sol Llama (Cloud-based)}}   & \sepehrgh{89.5} & \sepehrgh{92.9} & \sepehrgh{81.4} & \sepehrgh{100} \\
& \makecell{\sol OpenAI (Cloud-based)}            & 88.6 & 91.8 & 81.6 & 100 \\
& \makecell{\sol Lily (Local-based)}               & 63.8 & 63.4 & 75.1 & 63 \\
& \makecell{ChatGPT-off-the-shelf}                 & 14.8 & 25.8 & 71.1 & 87 \\
\hline

\multirow{4}{*}{\textbf{Avg.}} 
& \makecell{\sepehrgh{\sol Llama (Cloud-based)}}   & \sepehrgh{90.2} & \sepehrgh{91.7} & \sepehrgh{79.8} & \sepehrgh{100} \\
& \makecell{\sol OpenAI (Cloud-based)}            & 90.4 & 91.5 & 79.8 & 100 \\
& \makecell{\sol Lily (Local-based)}               & 60.7 & 65.9 & 75.0 & 71.7 \\
& \makecell{ChatGPT-off-the-shelf}                 & 20.1 & 25.1 & 70.6 & 92.3 \\
\hline
\end{tabular}
}
\end{table}

\end{document}